 \documentclass[12pt,3p]{elsarticle}
\biboptions{sort&compress}
\makeatletter
\def\ps@pprintTitle{%
  \let\@oddhead\@empty
  \let\@evenhead\@empty
  \def\@oddfoot{\reset@font\hfil\thepage\hfil}
  \let\@evenfoot\@oddfoot
}
\makeatother

\usepackage[utf8]{inputenc}
\usepackage[T2A,T1]{fontenc}
\usepackage[ukrainian,english]{babel}

\usepackage{amsmath,amsfonts,amssymb,amsthm}
\usepackage{bm}

\usepackage {graphicx}

\usepackage{hyperref}

\usepackage{array}
\usepackage{comment}

\allowdisplaybreaks[4]

\usepackage{color} 				
\definecolor{dgray}{rgb}{0.6,0.6,0.6}
\definecolor{dmag}{rgb}{0.6,0.0,0.6}
\definecolor{mbul}{rgb}{0.102, 0.42, 0.102} 
\usepackage[normalem]{ulem} 	

\begin{document}
\title{Self-consistent Hartree-Fock-Bogoliubov approach for bosons: self-eliminating divergence and pure pair condensate}
\author[1]{M.~Bulakhov}
\ead{bulakh@kipt.kharkov.ua}
\author[1]{A.S.~Peletminskii}
\ead{aspelet@kipt.kharkov.ua}
\affiliation[1]{
orgsnization={Akhiezer Institute for Theoretical Physics, National Science Center "Kharkiv Institute of Physics and Technology", NAS of Ukraine}, 
city={Kharkiv}, 
postcode={61108},  
country={Ukraine}}


\begin{abstract}
    We investigate the thermodynamic properties of an interacting Bose gas with a condensate within the energy-functional formulation of the Hartree-Fock-Bogoliubov (HFB) approach. For a contact interaction, we derive a self-consistent solution to the HFB equations that intrinsically eliminates divergence. This solution characterizes the equilibrium state featuring a condensate of correlated pairs of particles. We analyze the temperature dependence of key thermodynamic quantities such as condensate density, chemical potential, entropy, pressure, specific heat capacity at constant volume, and isothermal compressibility and compare them with predictions from the Popov approximation (PA). 
    We predict that the transition temperature shifts to higher values due interactions, with the HFB approach yielding a larger shift than the PA.
    Analysis of the compressibility indicates that a pure pair condensate is unstable, and the stable equilibrium corresponds to only a mixture of single-particle and pair condensates.
  \end{abstract}

\maketitle

\section{Introduction}
The phenomenon of Bose-Einstein condensation in an ideal gas has a long history, originating with the works of Bose \cite{Bose_1924_ZP} and Einstein \cite{Ein_SKPAW_1925} during the early stages of the development of quantum mechanics. The theoretical treatment of this phenomenon, even in the regime of weak interatomic interaction, proved to be a significant challenge. This is due to the well-known inapplicability of the thermodynamic perturbative approach, whose series expansion gives rise to divergent terms at zero momentum. The problem was solved by Bogoliubov for a homogeneous Bose gas, who developed a special technique based on replacing the creation and annihilation operators with zero momentum in the Hamiltonian by $c$-numbers \cite{Bogoliubov_JPhysUSSR_1947}. The Bogoliubov prescription for the field operators was employed to formulate the Gross-Pitaevskii approach \cite{Gross_NuovCim_1961,Pitaevskii_JETP_1961}, which became a powerful tool for investigating inhomogeneous condensates \cite{Pethick_2008,Pitaevskii_2016}. 
Although both approaches are simple and effective in describing key features of Bose-Einstein condensation, their validity is restricted to zero temperature and asymptotically weak interactions, whereas all experiments on Bose-Einstein condensates in dilute gases are necessarily conducted at finite temperatures \cite{Anderson_Science_1995,Davis_PRL_1995,Bradley_PRL_1995}.
In addition, a more thorough investigation of the ground state of a weakly interacting Bose gas within the Bogoliubov approach reveals the presence of correlated pairs of particles with opposite momentum
\cite{Bogolubov_2010,Akhiezer_1981}, which are analogous to Cooper pairs in the BCS theory \cite{Girardeau_PhysRev_1959,Valatin_PhysRev_1961,Nozieres_JPhysFr_1982}. Therefore, a central problem lies in developing theoretical frameworks that incorporate both interparticle interactions and finite-temperature effects on the condensate properties. 

For a complete description of Bose-Einstein condensates at finite temperatures, it is necessary to use advanced perturbative methods and self-consistent approximations \cite{ShiPhysRep1998,MorganJPhysB2000,Zagrebnov_PhysRep_2001,Proukakis_2008,Andersen_RevModPhys_2004}. In particular, an extension of the Bogoliubov theory to finite temperature, which formally leads to the elementary excitation spectrum at zero temperature but with a temperature-dependent condensate density, was developed by Popov \cite{Popov_1983,Popov_1987}. The Popov approximation (PA), which essentially ignores pair correlations, can be artificially derived from the more general Hartree–Fock–Bogoliubov (HFB) approach. 
In turn, the self-consistent HFB approach, in its various formulations \cite{Krasilnikov_PhysParNucl_1993,Akhiezer_PhysRep_1994,Griffin_PRB_1996, Zagrebnov_PhysRep_2001,Poluektov_LTP_2002,Proukakis_2008,Andersen_RevModPhys_2004}, includes not only normal (single-particle) averages but also anomalous averages
that describe both the single-particle condensate and the condensate of correlated pairs of particles. These pairs affect the thermodynamic properties of the system, the stability of the condensate, and the structure of the excitation spectrum. 
However, both the PA and the HFB approach are known to suffer from divergences in physical quantities at high momenta (ultraviolet divergence), resulting from the assumption of a contact interaction.
Generally, there are two ways to overcome the problem -- to employ the model potential, whose Fourier transform $U({\mathbf{p}})$ goes to zero for ${\mathbf{p}}\to \infty$ \cite{Romero-Rochin_2013_PLA,Bulakhov2018} or to renormalize the coupling constant associated with the $s$-wave scattering length \cite{Pethick_2008,Pitaevskii_2016}. 
In recent years, considerable attention has been devoted to the study of Bose–Einstein condensates at finite temperatures, as thermal effects play a crucial role in understanding the full thermodynamic behavior of such systems including the equation of state and the transition temperature shift due to interparticle interaction \cite{Mordini_2020_PRL,Ota_2020_PRA,Vianello_2024_ScientRep}.

The main purpose of this paper is to study finite-temperature effects in a weakly interacting Bose gas within a self-consistent, divergent-free HFB approach that includes pair anomalous averages. To this end, in Section~2, we present the energy-functional formulation of the HFB approach developed in \cite{Krasilnikov_PhysParNucl_1993,Akhiezer_PhysRep_1994}. Section~3 is devoted to deriving a system of coupled equations for the energy functional corresponding to the pairwise interaction Hamiltonian. 
In Section 4, for a contact interaction potential, we obtain a solution with a self-eliminating divergence (SSED), which allows us to avoid procedures such as cut-off of integrals, the use of model potentials or renormalization. 
This solution corresponds to a condensate of correlated pairs, commonly referred to as a pair condensate.
We also discuss some physical implications of the derived solution, such as zero pressure at zero temperature and the interpretation of the divergent term as an effective attraction.
Section 5 focuses on the investigation of the temperature dependence of key thermodynamic quantities, such as condensate density, chemical potential, pressure, entropy, specific heat capacity, and isothermal compressibility within SSED and PA.
In particular, we suggest a first-order phase transition between the normal state and the  state with broken U(1) symmetry. We also observe an increase in the transition temperature compared to that of an ideal gas.
We find that the isothermal compressibility indicates the instability of a pure pair condensate (or SSED), allowing us to conclude that the stable equilibrium corresponds to a mixture of single-particle and pair condensates.
Finally, Section~6, summarizes and discusses the results.

\section{Self-consistent formalism and basic equations}

Consider a many-body system of identical, spinless bosonic atoms of mass $m$ interacting via a pairwise potential $V(\mathbf{x})$. In momentum representation, the corresponding second-quantized Hamiltonian is given by
\begin{equation}
    H
    =
    \sum_{\mathbf{p}}
    \frac{p^{2}}{2m}
    a^{\dag}_{\mathbf{p}}a_{\mathbf{p}}
    +
    \frac{1}{2V}\sum_{\mathbf{p}_{1}\dots\mathbf{p}_{4}}U(\mathbf{p}_{1}
-
    \mathbf{p}_{3})a^{\dag}_{\mathbf{p}_{1}}a^{\dag}_{\mathbf{p}_{2}}a_{\mathbf{p}_{3}}a_{\mathbf{p}_{4}}
\delta_{\mathbf{p}_{1}+\mathbf{p}_{2},\mathbf{p}_{3}+\mathbf{p}_{4}}
,
\label{eq:Hamil}
\end{equation}
where $a^{\dagger}_{\mathbf{p}}$ and $a_{\mathbf{p}}$ are the creation and annihilation operators of bosonic atom in the state with momentum $\mathbf{p}$, $V$ is the volume of the system and 
\begin{equation}
    U({\mathbf{p}})
    =
    \int{d\mathbf{x}} V(\mathbf{x}) e^{-i\mathbf{px}/\hbar}.
\end{equation}
A direct application of this Hamiltonian to study equilibrium and non-equilibrium properties of many-body systems leads to significant and often insurmountable difficulties, already in the case of a normal state. Therefore, some approximate formulations, such as the thermodynamic perturbative approach or mean-field theories, need to be developed for the Hamiltonian under consideration \cite{Akhiezer_1981,Abrikosov_1963}. In turn, for systems with broken symmetries, the emergence of additional order parameters further complicates their description, even in the regime of weak interatomic interactions. 

The quasiparticle approach has proven to be a powerful framework for studying interacting systems with spontaneously broken symmetries. It is precisely this non-perturbative approach that we use to formulate the basic equations describing a superfluid Bose system with two distinct types of condensates. Following \cite{Krasilnikov_PhysParNucl_1993,Akhiezer_PhysRep_1994}, we introduce the general form of the statistical operator of an ideal non-equilibrium Bose gas with broken U(1) symmetry:
\begin{equation}
    \rho=\exp(Z-F)
    , \quad
    F
    =   
    a^{\dagger}_{\mathbf{p}_{1}}
    A_{\mathbf{p_{1}p}_{2}}
    a_{\mathbf{p}_{2}}
    +
    \frac12
    (
        a_{\mathbf{p}_{1}}
        B_{\mathbf{p}_{1}\mathbf{p}_{2}}
        a_{\mathbf{p}_{2}}
        +
        a^{\dagger}_{\mathbf{p}_{1}}
        B^{*}_{\mathbf{p}_{1}\mathbf{p}_{2}}
        a^{\dagger}_{\mathbf{p}_{2}}
    )
    +
    a^{\dagger}_{\mathbf{p}}
    C_{\mathbf{p}}
    +
    C^{*}_{\mathbf{p}}
    a_{\mathbf{p}}
    ,
\label{eq:StatOp}
\end{equation}
where we assume summation over the repeated indices associated with the momentum, unless otherwise specified. The unknown quantities $A_{\mathbf{p}_{1}\mathbf{p}_{2}}$ and $B_{\mathbf{p}_{1}\mathbf{p}_{2}}$ can be determined in equilibrium via a self-consistency procedure described below. This statistical operator modifies the standard Wick’s rule, leading to the appearance of both normal and anomalous averages. In particular, noting that following quantities are different from zero:
\begin{equation}
    {\rm Tr\,\rho a_{\mathbf{p}}}=b_{\mathbf{p}}, 
    \quad
    {\rm Tr\,\rho a^{\dagger}_{\mathbf{p}}}=b^{*}_{\mathbf{p}},
\end{equation}
we can perform a unitary $c$-number shift transformation 
\begin{equation}\label{eq:CShift}
Ua_{\mathbf{p}}U^{\dag}=a_{\mathbf{p}}+b_{\mathbf{p}}, \quad Ua^{\dag}_{\mathbf{p}}U^{\dag}=a^{\dag}_{\mathbf{p}}+b^{*}_{\mathbf{p}}
\end{equation}
to eliminate the linear terms associated with $C_{\mathbf{p}}$ in the statistical operator \eqref{eq:StatOp}. The quantity $Z$ is determined from the normalization condition $\textrm{Tr}\,\rho=1$. 
Thus, as a result of the unitary transformation \eqref{eq:CShift}, we have
\begin{gather}
    \textrm{Tr}\,\rho a^{\dag}_{\mathbf{p}_{1}}a_{\mathbf{p}_{2}}
    =
    f_{\mathbf{p}_{2}\mathbf{p}_{1}}
    +
    b^{*}_{\mathbf{p}_{1}}b_{\mathbf{p}_{2}},
\\
    \textrm{Tr}\,\rho a_{\mathbf{p}_{1}}a_{\mathbf{p}_{2}}
    =
    g_{\mathbf{p}_{2}\mathbf{p}_{1}}
    +
    b_{\mathbf{p}_{1}}b_{\mathbf{p}_{2}},
    \quad
    \textrm{Tr}\,\rho a^{\dagger}_{\mathbf{p}_{1}}a^{\dagger}_{\mathbf{p}_{2}}
    =
    g^{\dagger}_{\mathbf{p}_{2}\mathbf{p}_{1}}
    +
    b^{*}_{\mathbf{p}_{1}}b^{*}_{\mathbf{p}_{2}},
\end{gather}
where
\begin{equation}
    f_{\mathbf{p}_{2}\mathbf{p}_{1}}
    =
    \textrm{Tr}\,\rho^{(0)}a^{\dagger}_{\mathbf{p}_{1}}a_{\mathbf{p}_{2}}, 
    \quad 
    g_{\mathbf{p}_{2}\mathbf{p}_{1}}
    =
    \textrm{Tr}\,\rho^{(0)}a_{\mathbf{p}_{1}}a_{\mathbf{p}_{2}}
    , \quad 
    g^{\dagger}_{\mathbf{p}_{2}\mathbf{p}_{1}}
    =
    \textrm{Tr}\,\rho^{(0)}a^{\dagger}_{\mathbf{p}_{1}}a^{\dagger}_{\mathbf{p}_{2}}
    \label{eq:CorrF}
\end{equation}
and
$
    \rho^{(0)}
    =
    U\rho U^{\dagger}
    .
$ 
Consequently, the non-equilibrium state of the system under consideration is specified by the condensate amplitudes $b_{\mathbf{p}}$ and $b^{*}_{\mathbf{p}}$, as well as by the normal $f_{\mathbf{p}_{2}\mathbf{p}_{1}}$ ($f=f^{\dagger}$) and the anomalous  $g_{\mathbf{p}_{2}\mathbf{p}_{1}}$ and $g^{\dagger}_{\mathbf{p}_{2}\mathbf{p}_{1}}$ ($g=\tilde{g}$) distribution functions, where the tilde symbol denotes transposition. The condensate amplitudes describe a single-particle (Bogoliubov) condensate, while the presence of pair anomalous averages may signal the formation of a condensate of correlated pairs. It should be noted that the distinction between a condensate of correlated pairs and a molecular condensate lies in the nature of the interparticle binding. In molecules, particles form tightly two-body bound states, which then condense as single objects \cite{Timmermans_PhysRep_1999,Radzihovsky_AnnPhys_2008,Peletminskii_LTP_2014,Peletminskii_JPhysB_2017}. 
In contrast, correlated pairs, arising from interparticle interactions and quantum fluctuations in equilibrium, are neither localized nor long-lived in the sense of conventional molecules. Moreover, for a system with $N$ particles, the number of diatomic molecules is $N/2$, while the number of correlated pairs is $N(N-1)/2$.  
Therefore, the correlated pairs under consideration are similar to Cooper pairs in BCS theory without well-defined bound energy.
In Fermi systems, a smooth transition from a condensate of molecules to a condensate of delocolized Cooper pairs, occurring as the attraction between particles is varied, is known as the BEC-BCS crossover \cite{Zwerger_RevModPhys_2008,Zwerger_2011,Levin_RevModPhys_2024}.      
For further analysis, it is convenient to combine the normal and anomalous distribution functions into a block matrix and the condensate amplitudes into a column vector: 
\begin{equation}
\hat{f}=\left( \begin{array}{cc}
f & -g \\ g^{\dagger} & -1-\tilde{f}
\end{array}
\right), 
\quad
    \hat{b}
    =
    \left(
    \begin{array}{c}
    b
    \\
    b^{*}
    \end{array}
    \right)
    .
\end{equation}
According to \eqref{eq:CorrF}, the matrices $A$, $B$, and $B^{\dagger}$ that enter the statistical operator $\rho^{(0)}$ can be expressed in terms of the quantities $f$, $g$, and $g^{\dagger}$. Therefore, $\rho^{(0)}=\rho^{(0)}(\hat{f})$ and, consequently, the non-equilibrium entropy $S=-\textrm{Tr}\,\rho^{(0)}\ln\rho^{(0)}$ is also a certain functional of the normal and anomalous distribution functions, but not of the condensate amplitudes, $S=S(\hat{f})$. 
Note that $\hat{f}$ is isomorphic under a canonical transformation, which simultaneously reduces both $\rho^{(0)}$ and $\hat{f}$ to a diagonal form. 
As a result, one can show that (for details, see Ref.~\cite{Krasilnikov_PhysParNucl_1993,Akhiezer_PhysRep_1994}):
\begin{equation}
    \frac{\delta S(\hat{f})}{\delta f_{\mathbf{p}_{2}\mathbf{p}_{1}}}
    =
    A_{\mathbf{p}_{1}\mathbf{p}_{2}}, 
    \quad
    \frac{\delta S(\hat{f})}{\delta g_{\mathbf{p}_{2}\mathbf{p}_{1}}}
    =
    \frac{1}{2}
    B_{\mathbf{p}_{1}\mathbf{p}_{2}},
    \quad
    \frac{\delta S(\hat{f})}{\delta g^{\dagger}_{\mathbf{p}_{2}\mathbf{p}_{1}}}
    =
   \frac{1}{2}
    B^{\dagger}_{\mathbf{p}_{1}\mathbf{p}_{2}}.
\end{equation}
These formulas enable a self-consistency procedure by which the unknown quantities $A_{\mathbf{p}_{1}\mathbf{p}_{2}}$, $B_{\mathbf{p}_{1}\mathbf{p}_{2}}$, and $B^{\dagger}_{\mathbf{p}_{1}\mathbf{p}_{2}}$ are related to the physical characteristics of the system in equilibrium, which is determined by the principle of maximum entropy.

In order to formulate the variational principle for deriving the self-consistency equations governing the equilibrium state of the system, we note that the statistical operator \eqref{eq:StatOp} yields the following expectation values for the Hamiltonian operator $H$ and the particle number operator $N=\sum_{\mathbf{p}}a^{\dagger}_{\mathbf{p}}a_{\mathbf{p}}$:     
\begin{gather}
    E(\hat{f},\hat{b})
    = \textrm{Tr}\,\rho H, 
    \quad
    N(f,\hat{b})
=
\textrm{Tr}\,\rho N,
    \label{eq:ExpVal}
\end{gather}
where $N(f,\hat{b})$ is independent of anomalous distribution functions.
Then, the variational principle states that the equilibrium values of $\hat{f}$ and $\hat{b}$ should be found from the requirement of maximum entropy at fixed system energy and particle number. Therefore, introducing the Lagrange multipliers $\beta$ and $\mu$, we arrive at the following functional to be minimized:
\begin{equation}
 \mathcal{L}(\hat{f},\hat{b})
 =
 -S(\hat{f})
 +
 \beta E(\hat{f},\hat{b})
 -
 \mu N(f,\hat{b})
 .
\end{equation}
The solution of the formulated variational problem gives the following self-consistency equations: \cite{Krasilnikov_PhysParNucl_1993,Akhiezer_PhysRep_1994}:
\begin{gather}
\hat{f}=[\exp(\beta\hat{\xi})-1]^{-1}, \label{eq:BoseEq}
\\
\hat{\eta}-\mu\hat{b}=0,
\label{eq:GrossEq} 
\end{gather}
where
\begin{equation}
    \hat{\xi}
    =
    \left(
        \begin{array}{cc}
            \varepsilon-\mu & \Delta \\ 
            -\Delta^{*}     & -\tilde{\varepsilon}+\mu
        \end{array}
    \right)
    , 
    \quad 
    \hat{\eta}
    =
    \left(
    \begin{array}{c}
    \eta
    \\
    \eta^{*}
    \end{array}
    \right)
    \label{eq:QuasOper}
\end{equation}
and
\begin{equation}
    \varepsilon_{\mathbf{p}_{1}\mathbf{p}_{2}}
    =
    \frac{
        \delta E(\hat{f},\hat{b})
    }{
        \delta f_{\mathbf{p}_{2}\mathbf{p}_{1}}
    }
    , \quad 
    \Delta_{\mathbf{p}_{1}\mathbf{p}_{2}}
    =
    2
    \frac{
        \delta E(\hat{f},\hat{b})
    }{
        \delta g_{\mathbf{p}_{2}\mathbf{p}_{1}}
    }
    ,\quad
    \eta_{\mathbf{p}}
    =
    \frac{\delta E(\hat{f},\hat{b})}{\delta b^{*}_{\mathbf{p}}}.
    \label{eq:Notions}
\end{equation}
The derived equations determine the equilibrium values of the normal and anomalous distribution functions, as well as the condensate amplitude, for a given energy functional. Due to the bosonic structure of the operator $\hat{f}$ (see Eq.~\eqref{eq:BoseEq}), 
the Lagrange multipliers $\mu$ and $\beta=1/T$ should be interpreted as the chemical potential and the inverse temperature, respectively.
Equation \eqref{eq:GrossEq} is similar to the stationary Gross-Pitaevskii equation without external potential. However, the key distinction lies in the fact that the system’s energy now depends not only on the condensate amplitudes but also on the normal and anomalous distribution functions. It is worth noting that the quasiparticle formulation presented here is essentially equivalent to the HFB approach for bosons \cite{Andersen_RevModPhys_2004,Poluektov_LTP_2002}. Moreover, it is constructed in close analogy to the Fermi liquid theory extended to superfluid states \cite{Krasilnikov_PhysA_1990}.

\section{Energy functional and coupled equations for homogeneous state}

To make the coupled equations \eqref{eq:BoseEq}--\eqref{eq:Notions} suitable for further analysis, the system's energy functional must be specifed. According to \eqref{eq:Hamil}, \eqref{eq:ExpVal}, it is given by   
\begin{gather}
    E(\hat{f},\hat{b})
    =
    \sum_{\mathbf{p}}
    \frac{p^{2}}{2m}
    b^{*}_{\mathbf{p}}
    b_{\mathbf{p}}
    +
    \frac{1}{2V}
    \sum_{\mathbf{p}_{1}\dots\mathbf{p}_{4}}
    U(\mathbf{p}_{1}-\mathbf{p}_{3})
    b^{*}_{\mathbf{p}_{1}}
    b^{*}_{\mathbf{p}_{2}}
    b_{\mathbf{p}_{3}}
    b_{\mathbf{p}_{4}}
    \delta_{\mathbf{p}_{1}+\mathbf{p}_{2},\mathbf{p}_{3}+\mathbf{p}_{4}}
    \nonumber
    \\
    +
    \sum_{\mathbf{p}_{1}\mathbf{p}_{2}}
    \frac{p_{1}^{2}}{2m}
    f_{\mathbf{p}_{1}\mathbf{p}_{2}}
    \delta_{\mathbf{p}_{1},\mathbf{p}_{2}} 
    +
    \frac{1}{V}
    \sum_{\mathbf{p}_{1}\dots\mathbf{p}_{4}}f_{\mathbf{p}_{1}\mathbf{p}_{2}}
    \left[
        U(\mathbf{p}_{1}-{\bf p }_{2})+U(\mathbf{p}_{2}-\mathbf{p}_{4})
    \right] 
    b^{*}_{\mathbf{p}_{3}}
    b_{\mathbf{p}_{4}}
    \delta_{\mathbf{p}_{3}+\mathbf{p}_{2},\mathbf{p}_{1}+\mathbf{p}_{4}}  
    \nonumber \\
    +
    \frac{1}{2V}
    \sum_{\mathbf{p}_{1}\dots\mathbf{p}_{4}}
    U(\mathbf{p}_{2}-\mathbf{p}_{3})
    \left(
        g_{\mathbf{p}_{1}\mathbf{p}_{2}}
        b^{*}_{\mathbf{p}_{3}}
        b^{*}_{\mathbf{p}_{4}}
        +
        \mathrm{h.c.}
    \right)
    \delta_{\mathbf{p}_{1}+\mathbf{p}_{2},\mathbf{p}_{3}+\mathbf{p}_{4}}
    \nonumber \\
    +
    \frac{1}{2V}
    \sum_{\mathbf{p}_{1}\dots\mathbf{p}_{4}}
    f_{\mathbf{p}_{1}\mathbf{p}_{2}}
    f_{\mathbf{p}_{3}\mathbf{p}_{4}}
    \left[
        U(\mathbf{p}_{1}-\mathbf{p}_{2})
        +
        U(\mathbf{p}_{2}-\mathbf{p}_{3})
    \right]
    \delta_{\mathbf{p}_{1}+\mathbf{p}_{3},\mathbf{p}_{2}+\mathbf{p}_{4}} 
    \nonumber \\
    +
    \frac{1}{4V}
    \sum_{\mathbf{p}_{1}\dots\mathbf{p}_{4}}
    g^{\dagger}_{\mathbf{p}_{1}\mathbf{p}_{2}}
    g_{\mathbf{p}_{3}\mathbf{p}_{4}}
    \left[
        U(\mathbf{p}_{1}-\mathbf{p}_{3})
        +
        U(\mathbf{p}_{1}-\mathbf{p}_{4})
    \right]
    \delta_{\mathbf{p}_{1}+\mathbf{p}_{2},\mathbf{p}_{3}+\mathbf{p}_{4}}
    .
    \label{eq:EnerFunc}
\end{gather}
The expectation value of the particle number operator (see \eqref{eq:ExpVal}) reads
\begin{equation}
   N(f,\hat{b})
   =
   \sum_{\mathbf{p}_{1},\mathbf{p}_{2}}
   \left(
        f_{\mathbf{p}_1\mathbf{p}_2}
        +
        b^{*}_{\mathbf{p}_1}
        b_{\mathbf{p}_2}
    \right)
    \delta_{\mathbf{p}_1,\mathbf{p}_2}
    .
    \label{eq:TotPart}
\end{equation}

For a spatially homogeneous system, the coupled equations are considerably simplified. Indeed, in this case, we have
\begin{equation}
    f_{\mathbf{p}_1\mathbf{p}_2}
    =
    f_{\mathbf{p}_1}
    \delta_{\mathbf{p}_1\mathbf{p}_2}
    ,\quad
    g_{\mathbf{p}_1\mathbf{p}_2}
    =
    g_{\mathbf{p}_1}
    \delta_{\mathbf{p}_1,-\mathbf{p}_2}
    ,\quad
    b_{\mathbf{p}}
    =
    b_{0}
    \delta_{\mathbf{p},0}
    ,
    \label{eq:homogeneity1}
\end{equation}
where $f_{\mathbf{p}}=f^{*}_{\mathbf{p}}$ and $g_{\mathbf{p}}=g_{-\mathbf{p}}$. The above relations show that atoms with zero momentum and their correlated pairs with total zero momentum form a condensate. Equations \eqref{eq:Notions} then reduce to
\begin{gather}
    \varepsilon_{\mathbf{p}_1\mathbf{p}_2}
    =
    \varepsilon_{\mathbf{p}_1}\delta_{\mathbf{p}_1,\mathbf{p}_2}
    ,\quad 
    \Delta_{\mathbf{p}_1\mathbf{p}_2}
    =
    \Delta_{\mathbf{p}_1}\delta_{\mathbf{p}_1,\mathbf{p}_2}
    , 
    \label{eq:EDHom}
    \\
    \varepsilon_{\mathbf{p}}
    =
    \frac{\delta E(\hat{f},\hat{b})}
    {\delta f_{\mathbf{p}}}
    ,\quad 
    \Delta_{\mathbf{p}}
    =
    2
    \frac{\delta E(\hat{f},\hat{b})}
    {\delta g^{*}_{\mathbf{p}}}
    ,
    \label{eq:varder1}
    \\
    \mu b_{0}
    =
    \frac{
        \delta E(\hat{f},\hat{b})
    }{
        \delta b^{*}_{0}
    }
    .
    \label{eq:varder2}
\end{gather}
Performing simple algebraic transformations, one can show that equation \eqref{eq:BoseEq} yields
\begin{equation}
    f_{\mathbf{p}}
    =
    \frac{\xi_{\mathbf{p}}}
    {2E_{\mathbf{p}}}
    (1+2\nu_{\mathbf{p}})
    -
    \frac{1}{2}
    ,\quad
    g_{\mathbf{p}}
    =
    -
    \frac{\Delta_{\mathbf{p}}}
    {2E_{\mathbf{p}}}
    (1+2\nu_{\mathbf{p}})
    ,
    \label{eq:fg}
\end{equation}
where
\begin{equation}
    \nu_{{\mathbf{p}}}
    =[\exp(\beta E_{\mathbf{p}})-1]^{-1}
\end{equation}
and    
\begin{equation}
    \xi_{\mathbf{p}}
    =
    \varepsilon_{\mathbf{p}}
    -
    \mu, 
    \quad 
    E_{\mathbf{p}}
    =
    \sqrt{\xi^{2}_{\mathbf{p}}-|\Delta_{\mathbf{p}}|^{2}}
    .
    \label{eq:specdef}
\end{equation}

We now derive the coupled equations for the energy functional given by \eqref{eq:EnerFunc}. In this case, equation \eqref{eq:varder2} takes the form
\begin{gather}
    -
    b_{0}
    \mu
    +
    b_0
    n_{0}
    U(0)
    +
    \frac{b_0}{V}
    \sum_{\mathbf{p}'}
    f_{\mathbf{p}'}
    \left[
        U(0)
        +
        U(\mathbf{p}')
    \right]
    +
    \frac{b^{*}_0}{V}
    \sum_{\mathbf{p}'}
    g_{\mathbf{p}'}
    U(\mathbf{p}')
    =
    0,
    \label{eq:EqCondAmp}
\end{gather}
where $n_{0}=b^{*}_{0}b_{0}/V$ is the single-particle condensate density. Next, from equations \eqref{eq:varder1}, we find the quantities $\xi_{\mathbf{p}}$ and $\Delta_{\mathbf {p}}$ in terms of the distribution functions $f_{\mathbf{p}}$ and $g_{\mathbf{p}}$. Eliminating the latter (see \eqref{eq:fg}) from the corresponding expressions and equation \eqref{eq:EqCondAmp} and performing algebraic manipulations, we arrive at the following set of coupled equations for $\xi_{\mathbf{p}}$, $\Delta_{\mathbf{p}}$ and $b_{0}$ \cite{Tolmachev_1969,Peletminskii_2010,Poluektov_CMP_2013}:
\begin{gather}
    \xi_{\mathbf{p}}
    =
    \frac{p^{2}}{2m}
    -
    \mu
    +
    n_{0}\left[U(0)+U(\mathbf{p})\right]
    +
    \frac{1}{2V}
    \sum_{\mathbf{p}'}
    \left[
        U(0)
        +
        U(\mathbf{p}'-\mathbf{p})
    \right]
    \left(
        \frac{\xi_{\mathbf{p}'}}{E_{\mathbf{p}'}}
        \left(
            1+2\nu_{\mathbf{p}'}
        \right)
        -
        1
    \right)
    , \nonumber \\
    \Delta_{\mathbf{p}}
    =
    \frac{1}{V} 
    b^2_0
    U(\mathbf{p})
    -
    \frac{1}{2V}
    \sum_{\mathbf{p}'}
    U(\mathbf{p}+\mathbf{p}')
    \frac{\Delta_{\mathbf{p}'}}{E_{\mathbf{p}'}}
    \left(
        1+2\nu_{\mathbf{p}'}
    \right)
    , 
    \nonumber
     \\
    b_{0}
    \left(
    2n_{0}U(0)
    -
    \xi_{0}
    -
    \frac{b_{0}^{*}}{b_{0}}\Delta_{0}
    \right)
    =
    0. 
    \label{eq:xi,delta,b}
\end{gather}
The system \eqref{eq:xi,delta,b} should be supplemented by equation \eqref{eq:TotPart} for the total particle density $n=N/V$. In addition, we substitute $b_{0}=\sqrt{n_0V}e^{i\beta}$ and $\Delta_{\mathbf{p}}=|\Delta_{\mathbf{p}}|e^{i\gamma}$ into \eqref{eq:xi,delta,b} and separate the real and imaginary parts. As a result, we obtain
\begin{subequations}
\begin{equation}
    \xi_{\mathbf{p}} 
    =
    \frac{p^{2}}{2m}
    -
    \mu
    +
    n_{0}
    \left[
        U(0)+U(\mathbf{p})
    \right]
    +
    \frac{1}{2V}
    \sum_{\mathbf{p}'}
    \left[
        U(0)
        +
        U(\mathbf{p}'-\mathbf{p})
    \right]
    \left(
        \frac{\xi_{\mathbf{p}'}}{E_{\mathbf{p}'}}
        \left(
            1+2\nu_{\mathbf{p}'}
        \right)
        -
        1
    \right)
    , 
\end{equation}
\begin{equation}
    \Delta_{\mathbf{p}}
    =
    n_0
    U(\mathbf{p})
    \cos(\gamma-2\beta)
    -
    \frac{1}{2V}
    \sum_{\mathbf{p}'}
    U(\mathbf{p}+\mathbf{p}')
    \frac{
        \Delta_{\mathbf{p}'}
    }{
        E_{\mathbf{p}'}
    }
    \left(
        1+2\nu_{\mathbf{p}'}
    \right)
    , 
    \label{eq:Delta}
\end{equation}
\begin{equation}
    \left[
        2n_{0}U(0)
        -
        \xi_{0}
        -
        \Delta_{0}
        \cos(\gamma-2\beta)
    \right]
    \sqrt{n_{0}}
    =
    0
    ,
    \label{eq:chempot}
\end{equation}
\begin{equation}
    n
    =
    n_{0}
    +
    \frac{1}{2V}
    \sum_{\mathbf{p}'}
    \left(
        \frac{\xi_{\mathbf {p}'}}{E_{\mathbf{p}'}}(1+2\nu_{\mathbf{p}'})
        -
        1
    \right)
    ,
\end{equation}
\label{eq:CoupledSys}
\end{subequations}
where $\Delta_{\mathbf{p}}\equiv |\Delta_{\mathbf{p}}|\geq 0$ and
\begin{equation}   
    \cos(\gamma-2\beta)=\pm 1
    .
    \label{eq:cosConstrain}
\end{equation}
The coupled equations obtained, together with the definition of $E_{\mathbf{p}}$, allow us to find $n_{0}$, $\mu$, $\Delta_{\mathbf{p}}$ as functions of total density, temperature, and interaction parameters.

\section{Contact interaction}
\label{sec:contact}
It is clear that the system of equations \eqref{eq:CoupledSys} is difficult to solve, even using numerical methods, due to its transcendental and integral nature. Therefore, we should make a certain assumption that simplifies its analysis while still allowing us to obtain reasonable physical results. Specifically, we employ the fact that at low temperatures, low-energy collisions are characterized by the scattering length $a$. Therefore, we can replace the realistic interatomic potential with an effective contact pseudopotential that reproduces the correct scattering length \cite{Pitaevskii_2016}, 
\begin{equation}
    U=\frac{4\pi\hbar^{2}}{m} a.
\end{equation}
The coupled equations are now significantly simplified,
\begin{subequations}
\begin{equation}  
    \xi_{\mathbf{p}}
    =
    \frac{p^{2}}{2m}
    -
    \mu
    +
    2nU
    ,
\end{equation}
\begin{equation}
    \Delta
    =
    n_0
    U
    \cos(\gamma-2\beta)
    -
    \Delta U I
    , 
    \label{eq:DeltaU}
\end{equation}
\begin{equation}
    \left[
        \Delta
        \cos(\gamma-2\beta)
        +
        2(n-n_{0})U
        -
        \mu
    \right]
    \sqrt{n_{0}}
    =
    0
    ,
    \label{eq:chempotU}
\end{equation}
\begin{equation}
    n
    =
    n_{0}
    +
    \frac{1}{2V}
    \sum_{\mathbf{p}}
    \left(
        \frac{\xi_{\mathbf {p}}}{E_{\mathbf{p}}}(1+2\nu_{\mathbf{p}})
        -
        1
    \right)
    ,
    \label{eq:numbpartU}
\end{equation}
    \label{eq:CoupledSysU}
\end{subequations}
where
\begin{equation}
    I
    =
    \frac{1}{2V}
    \sum_{\mathbf{p}}
    \frac{
        1
    }{
        E_{\mathbf{p}}
    }
    \left(
        1+2\nu_{\mathbf{p}}
    \right)
    .
    \label{eq:Inf}
\end{equation}
Below, we are interested in the repulsive contact potential $U>0$. Therefore, since $\Delta>0$ and $I>0$, equations \eqref{eq:cosConstrain} and \eqref{eq:DeltaU} imply that $\cos(\gamma-2\beta) = 1$. In addition, it is straightforward to see that replacing $U(\mathbf{p})$ with the contact interaction $U$ leads to a divergent term $I$. This problem is commonly known and also arises in other approaches \cite{Pitaevskii_2016,Proukakis_2008,Andersen_RevModPhys_2004}. 
Several ways have been proposed to address this problem: replacing the upper limit of integration with a cut-off value, using a non-contact model potential, renormalizing the coupling constant, or simply neglecting the divergent term.

Imposing a cut-off at sufficiently large momentum $p_{c}$ or employing a model potential with a finite range $r_{0}$ to eliminate the divergence, are in fact, equivalent procedures. 
Metaphorically speaking, applying a cut-off to the integral is as different from using a model potential as emergency braking is from service braking.
In other words, the Fourier transforms of model potentials lead to a slow decay of the integrand, which ensures a convergence of the corresponding integrals \cite{Bulakhov2018}.
Nevertheless, a simple numerical analysis reveals that the solution to the equations is sensitive to small variations in $p_{c}$ or $r_{0}$, indicating that we cannot rely on the above methods to remove the divergence. 

As for renormalization, it is not applicable within the considered approach, as it leads to a complex energy spectrum $E_{\mathbf{p}}$. 
Indeed, in order to show this, avoiding complex calculations, note that renormalization can be effectively reduced to the following substitution in the equation \eqref{eq:DeltaU}:
\begin{equation}
    I
    \rightarrow
    \tilde{I}
    =
    \frac{1}{2V}
    \sum_{\mathbf{p}}
    \left(
        \frac{
            1
        }{
            E_{\mathbf{p}}
        }
        -
        \frac{
            2m
        }{
            p^2
        }
    \right)
    \left(
        1+2\nu_{\mathbf{p}}
    \right)
    <
    0
    .
\end{equation}
This results in $\Delta>n_0U$. Next, substituting $\Delta$ into \eqref{eq:specdef} and applying \eqref{eq:chempot}, we observe that even at ${\mathbf {p}} = 0$, the quantity
\begin{equation}   
    E^2_0
    =
    4n_0U
    (n_{0}U-\Delta)
    \label{eq:gap}
\end{equation}
is negative. 
Finally, other ways of dealing with the divergence prove to be more effective, so we will elaborate on them in greater detail below.

We now present the main thermodynamic quantities that we analyze in the next sections. The pressure is defined by the general formula:
\begin{equation}
    PV
    =
    TS-E+\mu N
    .
    \label{eq:pressure}
\end{equation}
Therefore, to compute it, we use the entropy expressed through the distribution function of an ideal gas of quasiparticles, according to the well-known formula
\begin{equation}
    S
    =
    k_{B}
    \sum_{\mathbf p}
    \left[
        (1+\nu_{\mathbf{p}})
        \ln(1+\nu_{\mathbf{p}})
        -
        \nu_{\mathbf{p}}
        \ln\nu_{\mathbf{p}}
    \right]
    .
    \label{eq:entropy}
\end{equation} 
Within the contact interaction approximation, the internal energy can be expressed in a relatively simple form by substituting equations \eqref{eq:fg} and \eqref{eq:CoupledSysU} into \eqref{eq:EnerFunc},
\begin{gather}
    E/V
    =
    \frac{1}{2V}
    \sum_{\mathbf{p}}
    \left(
        E_{\mathbf{p}}
        -
        \xi_{\mathbf{p}}
    \right)
    -
    (n^2-n_0^2)U
    -
    \frac{\Delta^2}{2U}
    +
    \frac{1}{V}
    \sum_{\mathbf{p}}
    E_{\mathbf{p}}
    \nu_{\mathbf{p}}
    +
    \mu n
    .
    \label{eq:homoEn}
\end{gather} 
Finally, from \eqref{eq:pressure} we obtain
\begin{gather}
    P
    =
    \frac{1}{2V}
    \sum_{\mathbf{p}}
    \left(
        \xi_{\mathbf{p}}
        -
        E_{\mathbf{p}}
    \right)
    +
    (n^2-n_0^2)U
    +
    \frac{\Delta^2}{2U}
    +
    \frac{k_{B}T}{V}
    \sum_{\mathbf{p}}
    \ln{(1+\nu_{\mathbf{p}})}
    .
    \label{eq:pressureU}
\end{gather}
We emphasize that equations \eqref{eq:entropy}--\eqref{eq:pressureU} are valid for both normal and degenerate states.

\subsection{Gapless spectrum and Popov approximation} \label{sec:gapless}

Any self-consistent solution of the coupled equations \eqref{eq:CoupledSysU} leads to a gapful spectrum of excitations. Nevertheless, let us demonstrate that neglecting the divergent term $I$ in the system \eqref{eq:CoupledSysU} yields a gapless Bogoliubov spectrum that satisfies the Pines–Hugenholtz condition \cite{Hugenholtz_PhysRev_1959}. It is worth noting that this approximation is artificial and does not correspond to a self-consistent solution of \eqref{eq:CoupledSysU}. Therefore, by neglecting the diverging term, one obtains
\begin{equation}
    \Delta^{\textrm{PA}}
    =
    n_{0}U.
\end{equation}
This immediately gives
\begin{equation}
    \xi^{\textrm{PA}}_{\mathbf{p}}
    =
    \frac{p^{2}}{2m}
    +
    n_0U
    ,
\end{equation}
\begin{equation}
    \mu^{\textrm{PA}}
    =
    n_0U
    +
    2(n-n_{0})U
    .
    \label{eq:chempotUgless}
\end{equation}
Substitution of these expression into \eqref{eq:specdef} results in the gapless Bogoliubov spectrum,
\begin{equation}
    E^{\textrm{PA}}_{\mathbf{p}}
    =
    \sqrt{
        \frac{p^{2}}{2m}
        \left(
            \frac{p^{2}}{2m}
            +
            2n_0U
        \right)
    }
    .
    \label{eq:specgless}
\end{equation}
Since $n_{0}=n_{0}(T)$, all relevant physical quantities implicitly depend on temperature. However, considering the zero temperature limit, $T=0$, and neglecting the depletion effect ($n=n_{0}$), we arrive at the well-known Bogoliubov relation $\mu\approx n_{0}U$. 

In this approximation, the pressure, according to \eqref{eq:pressure}, is given by
\begin{gather}
    P^{\textrm{PA}}
    =
    \frac{1}{2V}
    \sum_{\mathbf{p}}
    \left(
        \xi^{}_{\mathbf{p}}
        -
        E_{\mathbf{p}}
        -
        \frac{n_0^2U^2m}{p^2}
    \right)
    +
    n^2U
    -
    \frac{n_0^2U}{2}
    +
    \frac{k_{B}T}{V}
    \sum_{\mathbf{p}}
    \ln{(1+\nu_{\mathbf{p}})}
    ,
    \label{eq:pressurePH}
\end{gather}
where we performed the well-known regularization of the first term, which would diverge linearly at large momenta.
In particular, for $T=0$ ($n=n_{0}$), we obtain the Bogoliubov pressure, 
\begin{equation}
    P^{\textrm{PA}}\approx\frac{n^{2}_{0}U}{2}.
    \label{eq:pressureBog}
\end{equation}
The approximation presented here for $T\geq 0 $ is also known as the PA.

Finally, we note that the quantity $I$ determines the number of correlated particles (see \eqref{eq:Inf} and \eqref{eq:fg}). Neglecting this term amounts to disregarding the contribution of collective states, which has a significant impact on the structure of the single-particle spectrum \eqref{eq:specgless}, rendering it gapless (this issue is revisited in the final section).

\subsection{Solution with self-eliminating divergence}
\label{ssec:SEInf}

In this subsection, we propose an approach that enables the derivation of a finite and, more importantly, self-consistent solution to system \eqref{eq:CoupledSysU} without neglecting the divergence term. First and foremost, it is important to draw attention to the fact that may not be immediately apparent: the quantity $I$ diverges irrespective of the system’s parameters. As a result, equation \eqref{eq:DeltaU} is not mathematically constrained by the remaining equations of system \eqref{eq:CoupledSysU} and admits the following solution:
\begin{equation}
    \Delta^{\textrm{SSED}}
    =
    \lim_{I\rightarrow\infty}
    \frac{n_0U}{1+UI}
    =
    \lim_{I\rightarrow\infty}
    \frac{n_0}{I}
    =
    0
    .
    \label{eq:Deltagfull}
\end{equation}
Next, from \eqref{eq:chempotU}, we obtain the chemical potential:
\begin{equation}
    \mu^{\textrm{SSED}}
    =
    2 
    (n-n_0)
    U
    .
    \label{eq:chempotUgfull}
\end{equation}
Now the remaining equations from \eqref{eq:CoupledSysU} allow us to compute the energy spectrum, 
\begin{equation}
    E^{\textrm{SSED}}_{\mathbf{p}}
    =
    \xi^{\textrm{SSED}}_{\mathbf{p}}
    =
    \frac{p^2}{2m}
    +
    2
    n_0
    U
    \label{eq:specgfull}
\end{equation}
and the total particle density,
\begin{equation}
    n
    =
    n_0
    +
    \frac{1}{V}
    \sum_{\mathbf{p}}
    \nu^{\textrm{SSED}}_{\mathbf{p}}
    ,
    \quad
    \nu^{\textrm{SSED}}_{\mathbf{p}}
    =
    \left[
        \exp
        \left(
            \beta
            \xi^{\textrm{SSED}}_{\mathbf{p}}
        \right)
        -
        1
    \right]
    ^{-1}
    .
    \label{eq:specSSED}
\end{equation}
Note that at zero temperature, there is no depletion of the condensate ($n=n_{0}$), and the chemical potential vanishes, $\mu=0$. Let us now find the anomalous correlation function $g_{\mathbf{p}}$, which determines the number of paired particles. From \eqref{eq:fg}, we have:
\begin{equation}
    \frac{1}{V}\left|
    \sum_{\mathbf{p}}
    g_{\mathbf{p}}\right|
    =
    \lim_{I\rightarrow\infty}
    \frac{n_0}{I}
    \cdot 
    I
    =
    n_0
    .
    \label{eq:numpairSD}
\end{equation}
This implies that all correlated particles are in the condensate. At absolute zero temperature ($T=0$), all particles in the gas become simultaneously paired and condensed, giving rise to a macroscopically (correlated and coherent) quantum state (see \cite{Huang_1987}, section~2.3).

Thus, in the case of the described state, we should refer not to single-particle excitations, but to collective low-energy oscillations. 
Therefore, there is a connection between the structure of the spectrum and the density of the paired particles. Moreover, a remarkable point is that the energy gap \eqref{eq:gap} for the SSED analytically reaches a maximum value as a function of $\Delta$ for a given $n_{0}$.

In the case under consideration, the pressure takes the form:
\begin{equation}
     P^{\textrm{SSED}}
    =
    (n^2-n_0^2)U
    +
    \frac{k_{B}T}{V}
    \sum_{\mathbf{p}}
    \ln{(1+\nu_{\mathbf{p}})}
    .
    \label{eq:pressureSSED}
\end{equation}
From \eqref{eq:pressurePH}, it is evident that $P^{\textrm{PA}}>P^{\textrm{SSED}}$. Furthermore, we obtain the interesting result that $P^{\textrm{SSED}}=0$ precisely at $T=0$, as the divergence is naturally eliminated. To explain this outcome, we follow the reasoning provided by Huang in his discussion of the Van der Waals gas \cite{Huang_1987}, which served as an inspiration for understanding the vanishing pressure. Since the system of equations \eqref{eq:CoupledSysU} was derived  self-consistently, the divergent term $I$ can be interpreted as an effective attractive interaction between an individual pair of particles, as also indicated by the negative sign in equation  \eqref{eq:DeltaU}. The physical origin of this attraction lies in the cumulative repulsion from all other particles in the gas acting on the considered pair. 
Recalling that pressure is defined as force per unit area, we observe that in our SSED framework, repulsive and attractive forces are exactly balanced at every point. Furthermore, another argument supporting the interpretation of $I$ as an attraction comes from the very form of the Bogoliubov pressure \eqref{eq:pressureBog}. Within the Bogoliubov approximation, the contribution of $I$ must be neglected, resulting in a finite positive pressure that reflects the dominance of repulsive interactions in the gas. 

Let us emphasize that the obtained solution with a self-eliminating divergence arises due to the consideration of a contact potential (or pseudo-potential) with zero interaction range. For model potentials with a finite interaction range $r_{0}$, such a solution does not exist, since the term responsible for generating the quantity $I$ becomes finite. Indeed, from equations \eqref{eq:Delta} and \eqref{eq:DeltaU}, we have
\begin{equation}
    \lim_{r_0\rightarrow 0}U(\mathbf{p})
    =
    U
    \implies
    \lim_{r_0\rightarrow 0}
    \Delta(\mathbf{p})
    =
    \lim_{r_0\rightarrow 0}
    \frac{n_0}{I}
    =
    0
    \ 
    \land
    \
    \lim_{r_0\rightarrow 0}
    I
    =
    \infty
    .
\end{equation}
In other words, the limits $I\to \infty$ and $r_{0}\to 0$ are equivalent.

\subsection{Normal state}

In the preceding section, we presented the temperature-dependent solutions that describe only the degenerate state of the Bose gas, and not the normal state, where $n_{0}=0$. This can be proved by setting $n_{0}=0$ in the spectra \eqref{eq:specgless} and \eqref{eq:specgfull}, and verifying whether the conservation law for the total particle number, as given in \eqref{eq:numbpartU}, is satisfied. As we can see, it is not satisfied in either case, because the chemical potentials do not explicitly appear in the corresponding distribution functions $\nu_{\mathbf{p}}$, but are instead defined through their own dependencies, see \eqref{eq:chempotUgless} and \eqref{eq:chempotUgfull} at $n_{0}=0$.    

However, upon revisiting equation \eqref{eq:chempotU}, we observe that the left-hand side contains a factor of $\sqrt{n_{0}}$, which evidently vanishes in the absence of a condensate (normal state). Thus, this equation is automatically satisfied, and there is no need to determine the chemical potential in the manner used in the preceding subsections. This leads us to another self-consistent solution of the system \eqref{eq:CoupledSysU}, which we present omitting the algebraic derivation:
    \begin{gather}
        \Delta^{\textrm{NS}}
        =
        0
        ,\qquad
        E^{\textrm{NS}}_{\mathbf{p}}
        =
        \xi^{\textrm{NS}}_{\mathbf{p}}
        =
        \frac{p^2}{2m}
        -
        \mu
        +
        2nU
        ,
        \\
        n
        =
        \frac{1}{V}
        \sum_{\mathbf{p}}
        \nu^{\textrm{NS}}_{\mathbf{p}}
        ,
        \quad
        \nu^{\textrm{NS}}_{\mathbf{p}}
        =
        \left[
            \exp
            \left(
                \beta
                \xi^{\textrm{NS}}_{\mathbf{p}}
            \right)
            -
            1
        \right]
        ^{-1}
        .
        \label{eq:specNormstate}
    \end{gather}
This solution ensures the conservation law for the total particle number by appropriately adjusting the chemical potential $\mu$. Moreover, this state contains no information about the condensate.  Physically, this can be interpreted as the fact that, upon increasing (or decreasing) the temperature, the gas is compelled to undergo a phase transition to (or from) the normal state (described by \eqref{eq:specNormstate}) in the vicinity of some critical point.

The pressure in the normal state can be written as
\begin{gather}
    P^{\textrm{NS}}
    =
    n^{2}U
    +
    \frac{k_{B}T}{V}
    \sum_{\mathbf{p}}
    \ln{(1+\nu_{\mathbf{p}})}
    .
    \label{eq:pressureNS}
\end{gather}
In the ideal gas limit ($U=0$), the relations \eqref{eq:specNormstate} and \eqref{eq:pressureNS} completely reproduce the well-known expressions. Therefore, the terms $2nU$ in $\xi^{\textrm{NS}}_{\mathbf{p}}$ and $n^{2}U$ in $P^{\textrm{NS}}$ can be interpreted as interaction-induced corrections to the chemical potential and pressure, respectively, consistent with the results reported in \cite{Bulakhov_PS_2021,Ota_2020_PRA,Spada_PRA_2022}.

\section{Temperature dependence of physical observables}

Having obtained the solution to the system of equations \eqref{eq:CoupledSysU}, we now proceed to analyze the thermodynamic properties of a non-ideal Bose gas. Before doing so, let us briefly recall that for an ideal gas, the temperature dependence of the condensate density $n_{0}$ and the critical temperature $T^{\textrm{IBG}}$ are given by the following expressions:
\begin{equation*}
    n_{0}=n\left[1-\left(
    \frac{T}{T^{\textrm{IBG}}}
    \right)
    ^{3/2}\right]
    ,\qquad
    T^{\textrm{IBG}}
    =
    \frac{2\pi\hbar^2}{mk_B}
    \left(
        \frac{n}{\zeta(3/2)}
    \right)^{2/3}
    ,
\end{equation*}
where $k_{B}$ is the Boltzmann constant and $\zeta(x)$ is the Riemann zeta function. 

\begin{figure}[htb]
    \centering
    \includegraphics[width=81mm]{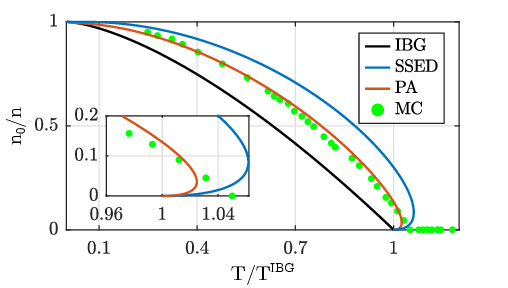}
    \caption{(Color online) Condensate fraction vs temperature for total density $n=1.6\cdot10^{14}$ cm$^{-3}$ and scattering length $a=4.9$ nm. The curves (from left to right) refer to: ideal Bose gas (IBG), Popov approximation (PA) and solution with self-eliminating divergence (SSED). 
    Magnified sections of the SSED and PA curves with retrograde behaviour are shown in the inset. The scatter plot (MC) illustrates the predictions of the universal relations, calculated using the Monte-Carlo method \cite{Prokof'ev_PRA_2004}.}
    
    \label{fig:dens}
\end{figure}

\begin{figure}[htb]
    \centering
    \includegraphics[width=81mm]{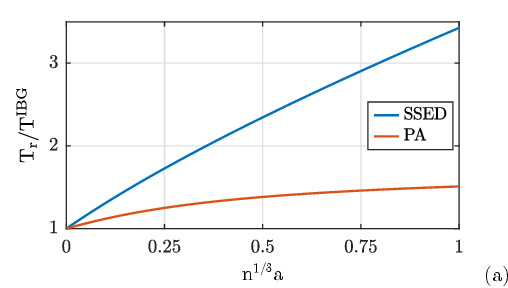}
    \includegraphics[width=81mm]{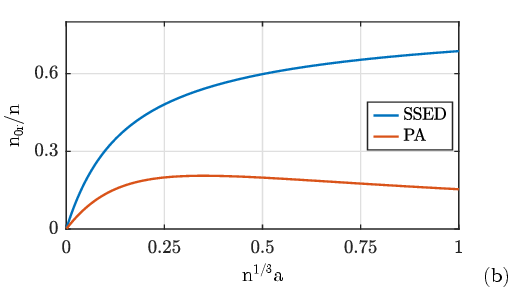}
    \caption{(Color online) Dependencies on gas parameter of: (a) reversal temperature, (b) condensate fraction at reversal temperature. The curves refer to: solution with self-eliminating divergence (SSED) and Popov approximation (PA).}
    \label{fig:reversalTemperature}
\end{figure}

Fig.~\ref{fig:dens} shows the temperature dependence of the condensate fraction for an ideal Bose gas, the PA and the SSED obtained from the coupled equations of the HFB approach. For comparison, Fig.~\ref{fig:dens} also displays Monte Carlo results taken from Ref.~\cite{Prokof'ev_PRA_2004}. The curves for the PA and SSED are obtained by numerically solving equation \eqref{eq:numbpartU}, which ensures the conservation of the total particle number. A notable feature of the presented dependencies is the retrograde behavior of the condensate density, which decreases to zero at $T=T^{\textrm{IBG}}$ following a reversal point at certain temperature $T=T_r$. While this behavior is qualitatively similar across different theoretical approaches (PA and SSED), the reversal temperature $T_{r}$ exhibits quantitative dependence on both the approximation used and the specific gas parameters. In particular, the dependencies of $T_{r}$ and the condensate fraction at this temperature on the gas parameter are demonstrated in Fig.~\ref{fig:reversalTemperature}. In the framework of PA, we observe a non-monotonic behavior of the condensate fraction on interaction strength, as also reported in \cite{Vianello_2024_ScientRep}. The nontrivial temperature dependence of the condensate fraction is clearly a consequence of the nonlinear structure of equation \eqref{eq:numbpartU}.  

\begin{figure}[tb]
    \centering
    \includegraphics[width=81mm]{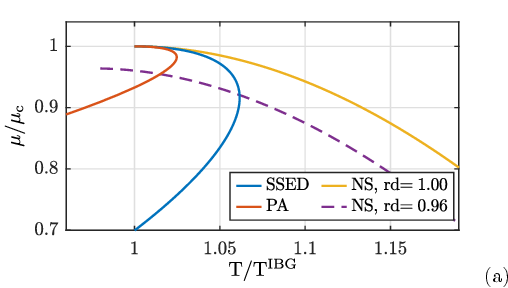}
    \includegraphics[width=81mm]{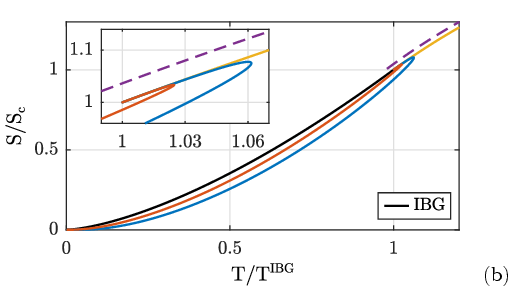}
    \includegraphics[width=81mm]{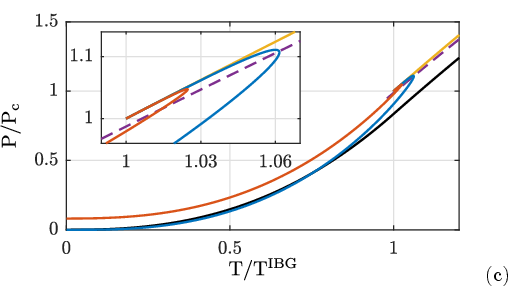}
    \includegraphics[width=81mm]{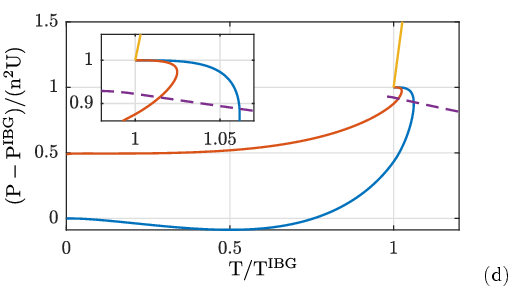}
    \caption{
        (Color online) Temperature dependencies of: (a) chemical potential; (b) entropy; (c) pressure; and (d) pressure relative to that of an ideal Bose gas --- for the total density of degenerate state $n^{\textrm{deg}}=1.6\cdot10^{14}$ cm$^{-3}$, the scattering length $a=4.9$ nm, $\mu_c=\mu^{\textrm{NS}}(T=T^{\textrm{IBG}})=2nU$, $S_c=S^{\textrm{NS}}(T=T^{\textrm{IBG}})=5/2\zeta(5/2)/\zeta(3/2)Nk_B$ and $P_c=P^{\textrm{NS}}(T=T^{\textrm{IBG}})=n^2U+\zeta(5/2)/\zeta(3/2)nk_BT^{\textrm{IBG}}$.
        The solid curves refer to: ideal Bose gas (IBG), solution with self-eliminating divergence (SSED), Popov approximation (PA) and normal state (NS) --- for total density $n=n^{\textrm{deg}}$, i.e. relative density rd$=n/n^{\textrm{deg}}=1$. The dashed curves refer to the normal state for rd$=0.96$.
        Magnified sections of the foregoing curves with retrograde behaviour are shown in the insets.
    }
    \label{fig:chem}
\end{figure}

The aforementioned retrograde temperature dependence is reflected in other thermodynamic quantities, including the chemical potential $\mu$, entropy $S$, and pressure $P$, as shown in Fig.~\ref{fig:chem}. 
The chemical potential of the normal state of the interacting Bose gas was computed numerically from \eqref{eq:specNormstate}. 
To plot the entropy, we employed Eq.~\eqref{eq:entropy} with the corresponding substitutions of \eqref{eq:specgless}, \eqref{eq:specSSED}, and \eqref{eq:specNormstate}. 
As for the pressure in the normal and degenerate states, we addressed the corresponding equations \eqref{eq:pressurePH}, \eqref{eq:pressureSSED}, and \eqref{eq:pressureNS}.
In both approaches (PA and SSED), the pressure of the interacting gas demonstrates a slight quantitative deviation from that of the ideal gas, while remaining positive throughout the entire temperature range. 
Moreover, the pressure difference between the interacting and ideal gases is smaller in the presence of a condensate than in the normal state.

Remarkably, despite the nonlinearity, chemical potentials and pressures of the degenerate and normal states coincide at $T=T^{\textrm{IBG}}$. 
Thus, the formal conditions for a phase transition are satisfied (see \cite{Huang_1987}). 
However, due to the retrograde behavior, this does not apply to the temperature derivatives of the foregoing quantities. 
Although the equality of derivatives at the transition point is not a necessary condition for the phase transition, the very existence of the derivatives is crucial. 
Indeed, the temperature derivative does not exist on the left side of the transition point, neither in the degenerate nor in the normal state. Therefore, the point $T=T^{\textrm{IBG}}$ cannot be identified as a genuine phase transition. Instead, the entire temperature interval $T\in[T^{\textrm{IBG}},T_{r}]$ should be considered as a region of instability.

For the phase transition to occur, it is sufficient that the right-hand derivative in the normal phase and left-hand derivative in the degenerate state exist. The temperature $T_{r}$ is the nearest point at which this condition may be fulfilled. The requirement of equal chemical potentials and pressures at this point necessitates adjusting the density of one of the phases. This situation is illustrated in Figs.~\ref{fig:chem} by a dashed line, where the density of the normal state is adjusted accordingly. Although the dashed line is shown only for the SSED case, it is evident that the same applies to the PA case.  

Since the total number of particles is conserved while the densities differ between phases, the phase transition is accompanied by discontinuities in volume and entropy, as shown in Fig.~3(b). As is well known, this behavior corresponds to a first order phase transition \cite{Huang_1987}. In particular, the entropy discontinuity leads to a nonzero latent heat. In addition, the order parameter also exhibits a discontinuous jump, as shown in Fig.~\ref{fig:dens}. Employing this fact, we can clarify the volume discontinuity from a physical standpoint. Since the condensate does not contribute to the pressure, and a transition occurs at constant pressure, the loss of the latter is compensated by a decrease in volume. 

\begin{figure}[htb]
    \centering
    \includegraphics[width=81mm]{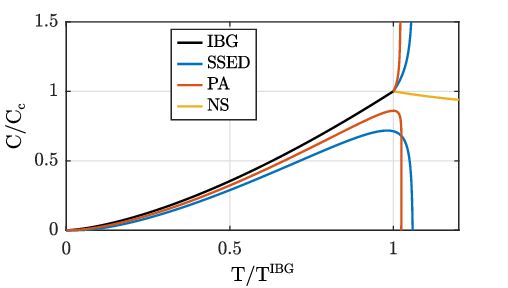}
    \caption{
        (Color online) Specific heat capacity at constant volume vs temperature for total density $n=1.6\cdot10^{14}$ cm$^{-3}$, scattering length $a=4.9$ nm and $C_c=C^{\textrm{NS}}(T=T^{\textrm{IBG}})=15/4\zeta(5/2)/\zeta(3/2)Nk_B$. The curves refer to: ideal Bose gas (IBG), solution with self-eliminating divergence (SSED), Popov approximation (PA) and normal state (NS). IBG and NS curves coincide for $T>T^{\textrm{IBG}}$.
    }
    \label{fig:shc}
\end{figure}

Considering the specific heat capacity at constant volume, we start by recalling
its definition: 
\begin{equation}
    C
    =
    \left(
    \frac{
        \partial 
        E
    }{
        \partial
        T
    }
    \right)_{V,N}
    =
    T
    \left(
    \frac{
        \partial 
        S
    }{
        \partial 
        T
    }
    \right)_{V,N}
    .
\end{equation}
As we can see in Fig.~\ref{fig:shc}, there is no crucial evidence (for example, $C>0$) pointing on any instability of the SSED compared to the PA. Neither the entropy nor the specific heat capacity in the normal state have linear corrections in interaction \cite{Akhiezer_1981,Bulakhov_PS_2021}. Therefore, these quantities coincide with those of the ideal Bose gas.

\begin{figure}[htb]
    \centering    \includegraphics[width=81mm]{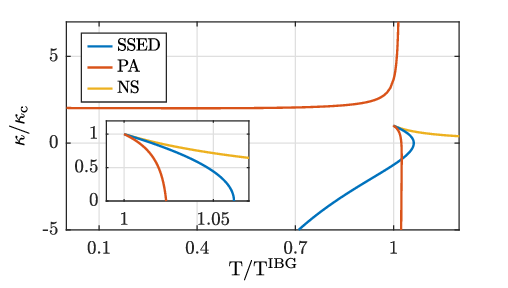}
    
    \caption{(Color online) Compressibility vs temperature for total density $n=1.6\cdot10^{14}$ cm$^{-3}$ and scattering length $a=4.9$ nm and $\kappa_c=\kappa^{\textrm{NS}}(T=T^{\textrm{IBG}})=(2n^2U)^{-1}$. The curves refer to: solution with self-eliminating divergence (SSED), Popov approximation (PA) and normal state (NS). Magnified sections of the foregoing curves with retrograde behaviour are shown in the inset.}
    \label{fig:compressibility}
\end{figure}

Another important thermodynamic quantity we wish to address is the isothermal compressibility, which plays a crucial role in characterizing the system's response to pressure changes at constant temperature,
\begin{equation}
    \kappa
    =
    \frac{1}{n}
    \left(
        \frac{\partial n}{\partial P}
    \right)_{T,N}.
\end{equation}
Its temperature dependence is presented in Fig.~\ref{fig:compressibility}, where the significant qualitative differences between the two approximations become apparent. The PA exhibits a positive and generally well-behaved compressibility (with divergence only in the region of instability), as also reported in \cite{Ota_2020_PRA,Mordini_2020_PRL,Spada_PRA_2022}. In contrast, for the SSED, the compressibility is negative almost everywhere, diverging to minus infinity as $T\to 0$.
Thus, the equilibrium state corresponding to the SSED is unstable.

To summarize, we observe that almost all thermodynamic quantities exhibit well-behaved properties within both the HFB approach and the PA. 
Furthermore, the latter, which accounts solely for the single-particle condensate, predicts the positive compressibility. However, since the HFB approach does not admit a solution of \eqref{eq:xi,delta,b} corresponding to a pure single-particle condensate \cite{Poluektov_LTP_2002,Poluektov_CMP_2013}, and the pair condensate (SSED) is unstable, the only possible stable state can be a mixture of single-particle and pair condensates , as also discussed below.

\section{Discussion of results}
\label{sec:discussion}
We have employed the energy-functional formulation 
to derive the self-consistency equations of the Hartree-Fock-Bogoliubov (HFB) approach. In the limit of contact interactions, these equations exhibit ultraviolet divergence, which are typically addressed by regularizing the corresponding integrals or renormalizing the coupling constant. We have obtained a new solution to the HFB self-consistency equations in which the divergence self-eliminates (the solution with self-eliminating divergence, SSED).
This solution possesses several distinct features. Specifically, it predicts the existence of a condensate in the form of correlated pairs only, with zero depletion and pressure at absolute zero temperature. The vanishing pressure at zero temperature can be explained by treating the divergent term as an effective attraction between an individual pair of particles, balanced by cumulative repulsion exerted by all other particles in the gas on the given pair. 
We have analyzed the temperature behavior of key thermodynamic quantities in the normal state and in the degenerate states described both by the obtained SSED and Popov approximation (PA). Fig.~\ref{fig:dens} shows a good agreement of the PA and Monte-Carlo. The respective comparison with experimental data presented in \ref{app:expt}.

The common feature of both mean-field approaches is the retrograde behaviour of the condensate density, which is also observed in Ref.~\cite{Prokof'ev_PRA_2004} for the PA, which in turn affects other thermodynamic quantities.
Nevertheless, this behaviour allows us to conclude that, due to interactions, the transition temperature shifts to higher values compared to that for an ideal gas, with the HFB approach yielding a larger shift than the PA. 
This positive shift of the transition temperature has also been reported in previous mean-field studies \cite{Vianello_2024_ScientRep,Arnold_PhysRevLett2001} and Monte Carlo simulations \cite{Prokof'ev_PRA_2004,Ota_2020_PRA,Spada_PRA_2022}. 
In both cases, we propose to treat the phase transition between the normal and degenerate state as a first-order one. 
Unfortunately, the isothermic compressibility for SSED appears to be negative, which signals its instability. 
We emphasize that the differences among all the considered approaches, including the ideal gas case, become more pronounced as the temperature approaches $T^{\textrm{IBG}}$, which facilitates experimental verification. 
At the same time, it should be noted that both approaches fail to provide an adequately accurate description of thermodynamics in the vicinity of the critical temperature $T^{\textrm{IBG}}$.
Therefore, an advanced possible theory that goes beyond the current mean-field approximations (PA and SSED) should not give rise to retrograde behavior and may change the
nature of the phase transition.

Some additional comments should also be made on certain important aspects of our study.
First, although model non-contact potentials are not explicitly considered in this paper, some conclusions can nevertheless be drawn regarding their influence on the obtained results. 
As noted in Section~\ref{sec:contact}, while they allow for a straightforward removal of divergences, the results lack stability with respect to variations in the interaction range and the choice of model potential. 
Furthermore, even within framework of Bogoliubov’s quadratic approximation, all solutions involving local potentials are also unstable in the same sense \cite{Bulakhov2018}. 
Within the HFB approach, as indicated by \eqref{eq:gap}, the gap lies in the range $E_0\in(0,2n_0U]$, since PA is gapless and SSED corresponds to $E_{0}=2n_{0}U$. 
This allows a qualitative conclusion that for non-contact potentials, the expected results will be bounded between the curves obtained from the SSED and the PA. 

In connection with the Nambu-Goldstone (NG) theorem \cite{Nambu_PhysRev_1960,Goldstone_NuovoCim_1961} and the Pines–Hugenholtz relation \cite{Hugenholtz_PhysRev_1959}, it is commonly believed that the presence of a gap in the single-particle excitation spectrum is a drawback of the self-consistent HFB approach. 
Numerous attempts have been made to artificially remove the gap in order to satisfy this relation. 
In particular, the most well-known gapless PA neglects anomalous averages, and in some literature sources \cite{Spada_PRA_2022}, the HFB approach is even considered to be valid only in the linear order in interaction. 
However, as we previously pointed out, these do not represent self-consistent solutions to the coupled HFB equations. 
We report a unique self-consistent solution to the HFB equations for a contact interaction that resolves the divergence issue without introducing additional assumptions or artificial regularization procedures. At sufficiently high temperatures, this solution exhibits qualitative agreement with the PA, except for the isothermal compressibility.

Therefore, the gathered evidence encourages us to reconsider the nature of the energy gap in the HFB approach. On the one hand, the presence of gapless (phonon) excitations is not disputed. 
On the other hand, the HFB approach, along with the PA artificially derived from it, represent an attempt to construct a quasiparticle description of an interacting system in terms of a single type of single-particle excitations. This holds true if the spectra of single-particle and collective excitations coincide \cite{GavoretAnnPhys1964}. 
However, it has been recently demonstrated that the poles of the single-particle and two-particle Green’s functions in a Bose system with condensate do not coincide \cite{Kita_PRB_2009,Kita_JPhysSocJap_2011}. 
Therefore, when both types of excitations predominate in the system, the application of the HFB framework leads to a unified spectrum of single-particle excitations hybridized with collective (pair) excitations, inheriting a gap from the latter. The presence of a gap in the two-particle excitations, however, is not an artifact, as it can be associated with the energy required to break a pair.

The NG theorem states that the excitation spectrum arising in a system with spontaneously broken symmetry must be gapless. 
This is equivalent to the statement that it should cost zero energy to excite the lowest mode (the Goldstone mode). 
While the direct application of the NG theorem to quasiparticles, which are not physical particles, remains debated \cite{Suto_PRA_2008}, we highlight the following observation. For the SSED, where $\Delta = 0$, the second equation in 
\eqref{eq:varder1} shows that the energy change associated with the emergence of a single correlation is zero. Therefore, when pair correlations dominate in the ground state, as described by the solution we have found, the lowest mode is collective, and its excitation costs no energy.

Thus, as we have already noted, a system of interacting particles in the presence of a pair condensate requires an advanced description, which, in our view, should also reproduce a spectrum with at least two excitation branches: a phonon-like mode and a gapped mode.
Similar conclusions have been drawn in Refs. \cite{Bogolubov_2010, Poluektov_JLTP_2023}.
It should be emphasized that pure single-particle or pair condensates do not exist -- only their mixture does. The existence of a purely single-particle condensate, as assumed in the PA, is ruled out by the system of HFB equations \eqref{eq:xi,delta,b} (see also \cite{Poluektov_LTP_2002, Poluektov_CMP_2013}). 
As for a pure pair condensate described by the SSED, it is found to be unstable based on the compressibility analysis. 
This result holds for the contact interaction model. In the case of finite-range interaction potentials, the corresponding HFB solutions also exhibit a gap, with $E_{0}<2n_{0}U$, and the equilibrium state likewise represents only a mixture of the two types of condensates.
Such a structure of the ground state is also relevant in the context of superfluid helium, where the phonon-roton spectrum emerges from the hybridization of excitations of different physical natures and the single-particle condensate fraction is about 10\% or less \cite{Sosnick_EPL_1989,Bogoyavlenskii_LTP_1990,Moroni_JLTP_2004,Glyde_PhysRevB_2011}.   

\section*{Acknowledgment}

M. Bulakhov acknowledges funding by the National Research Foundation of Ukraine, Grant No. 0124U004372. A.S. Peletminskii acknowledges support from the STCU project “Magnetism in Ukraine Initiative,” No. 9918.

\begin{appendix}
    \section{Comparison with experimental data}
    \label{app:expt}
\begin{figure}[h]
    \centering
    \includegraphics[width=81mm]{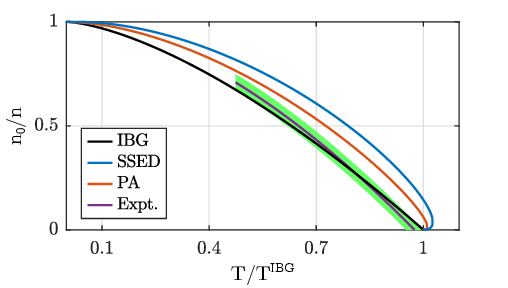}
    \caption{
        (Color online) Condensate fraction vs temperature for total density $n=1.24\cdot10^{13}$ cm$^{-3}$ and scattering length $a=4.9$ nm. The curves (from left to right) refer to: ideal Bose gas (IBG),
        experimental data  (Expt.) \cite{Gaunt_PhysRevA2013}, Popov approximation (PA) and 
        solution with self-eliminating divergence (SSED). The green band represents the experimental error.
    }
    \label{fig:comparisonToExpt}
\end{figure}

    Fig.~\ref{fig:comparisonToExpt} shows the temperature dependence of the condensate fraction, calculated within the mean-field approximations considered in this work for a uniform gas, compared with the experimental data of Ref.~\cite{Gaunt_PhysRevA2013}. These measurements were, according to the authors, performed under conditions of minimal trap influence and should therefore provide the most suitable benchmark. However, the necessary parameters for a quantitative comparison, such as density and scattering length, are not explicitly given by the authors. Therefore, the density was determined from the theoretically estimated critical temperature given by the authors, $T^{\textrm{IBG}}=98$ nK, while the scattering length was taken as a typical value for ${}^{87}$Rb atoms. A similar interpretation of the experimental data was performed in Ref.~\cite{Rakhimov_PhysLettA2025}. As seen in the figure, the accuracy of the mean-field approximations decreases with increasing temperature, confirming the conclusion of Section \ref{sec:discussion} that both mean-field and perturbative approaches fail in the vicinity of the phase transition.

\end{appendix}

\bibliographystyle{elsarticle-num}
\bibliography{main}

\begin{thebibliography}{62}
\expandafter\ifx\csname natexlab\endcsname\relax\def\natexlab#1{#1}\fi
\expandafter\ifx\csname bibnamefont\endcsname\relax
  \def\bibnamefont#1{#1}\fi
\expandafter\ifx\csname bibfnamefont\endcsname\relax
  \def\bibfnamefont#1{#1}\fi
\expandafter\ifx\csname citenamefont\endcsname\relax
  \def\citenamefont#1{#1}\fi
\expandafter\ifx\csname url\endcsname\relax
  \def\url#1{\texttt{#1}}\fi
\expandafter\ifx\csname urlprefix\endcsname\relax\def\urlprefix{URL }\fi
\providecommand{\bibinfo}[2]{#2}
\providecommand{\eprint}[2][]{\url{#2}}

\bibitem[{\citenamefont{Bose}(1924)}]{Bose_1924_ZP}
\bibinfo{author}{\bibfnamefont{S.~N.} \bibnamefont{Bose}}, \bibinfo{journal}{Z.
  Phys.} \textbf{\bibinfo{volume}{26}}, \bibinfo{pages}{178}
  (\bibinfo{year}{1924}).

\bibitem[{\citenamefont{Einstein}(1925)}]{Ein_SKPAW_1925}
\bibinfo{author}{\bibfnamefont{A.}~\bibnamefont{Einstein}},
  \bibinfo{journal}{Sitzungsber. Kgl. Preuss. Akad. Wiss.}
  \textbf{\bibinfo{volume}{1}}, \bibinfo{pages}{3} (\bibinfo{year}{1925}).

\bibitem[{\citenamefont{Bogoliubov}(1947)}]{Bogoliubov_JPhysUSSR_1947}
\bibinfo{author}{\bibfnamefont{N.}~\bibnamefont{Bogoliubov}},
  \bibinfo{journal}{J. Phys. USSR} \textbf{\bibinfo{volume}{11}},
  \bibinfo{pages}{23} (\bibinfo{year}{1947}).

\bibitem[{\citenamefont{Gross}(1961)}]{Gross_NuovCim_1961}
\bibinfo{author}{\bibfnamefont{E.~P.} \bibnamefont{Gross}},
  \bibinfo{journal}{Nuovo Cim. (1955-1965)} \textbf{\bibinfo{volume}{20}},
  \bibinfo{pages}{454} (\bibinfo{year}{1961}),
  \urlprefix\url{https://doi.org/10.1007/BF02731494}.

\bibitem[{\citenamefont{Pitaevskii}(1961)}]{Pitaevskii_JETP_1961}
\bibinfo{author}{\bibfnamefont{L.~P.} \bibnamefont{Pitaevskii}},
  \bibinfo{journal}{Sov. Phys. JETP} \textbf{\bibinfo{volume}{13}},
  \bibinfo{pages}{451} (\bibinfo{year}{1961}).

\bibitem[{\citenamefont{Pethick and Smith}(2008)}]{Pethick_2008}
\bibinfo{author}{\bibfnamefont{C.~J.} \bibnamefont{Pethick}} \bibnamefont{and}
  \bibinfo{author}{\bibfnamefont{H.}~\bibnamefont{Smith}},
  \emph{\bibinfo{title}{Bose--Einstein condensation in dilute gases}}
  (\bibinfo{publisher}{Cambridge university press}, \bibinfo{year}{2008}).

\bibitem[{\citenamefont{Pitaevskii and Stringari}(2016)}]{Pitaevskii_2016}
\bibinfo{author}{\bibfnamefont{L.}~\bibnamefont{Pitaevskii}} \bibnamefont{and}
  \bibinfo{author}{\bibfnamefont{S.}~\bibnamefont{Stringari}},
  \emph{\bibinfo{title}{Bose-Einstein condensation and superfluidity}}, vol.
  \bibinfo{volume}{164} (\bibinfo{publisher}{Oxford University Press},
  \bibinfo{year}{2016}).

\bibitem[{\citenamefont{Anderson et~al.}(1995)\citenamefont{Anderson, Ensher,
  Matthews, Wieman, and Cornell}}]{Anderson_Science_1995}
\bibinfo{author}{\bibfnamefont{M.~H.} \bibnamefont{Anderson}},
  \bibinfo{author}{\bibfnamefont{J.~R.} \bibnamefont{Ensher}},
  \bibinfo{author}{\bibfnamefont{M.~R.} \bibnamefont{Matthews}},
  \bibinfo{author}{\bibfnamefont{C.~E.} \bibnamefont{Wieman}},
  \bibnamefont{and} \bibinfo{author}{\bibfnamefont{E.~A.}
  \bibnamefont{Cornell}}, \bibinfo{journal}{Science}
  \textbf{\bibinfo{volume}{269}}, \bibinfo{pages}{198} (\bibinfo{year}{1995}),
  \urlprefix\url{https://www.science.org/doi/abs/10.1126/science.269.5221.198}.

\bibitem[{\citenamefont{Davis et~al.}(1995)\citenamefont{Davis, Mewes, Andrews,
  van Druten, Durfee, Kurn, and Ketterle}}]{Davis_PRL_1995}
\bibinfo{author}{\bibfnamefont{K.~B.} \bibnamefont{Davis}},
  \bibinfo{author}{\bibfnamefont{M.~O.} \bibnamefont{Mewes}},
  \bibinfo{author}{\bibfnamefont{M.~R.} \bibnamefont{Andrews}},
  \bibinfo{author}{\bibfnamefont{N.~J.} \bibnamefont{van Druten}},
  \bibinfo{author}{\bibfnamefont{D.~S.} \bibnamefont{Durfee}},
  \bibinfo{author}{\bibfnamefont{D.~M.} \bibnamefont{Kurn}}, \bibnamefont{and}
  \bibinfo{author}{\bibfnamefont{W.}~\bibnamefont{Ketterle}},
  \bibinfo{journal}{Phys. Rev. Lett.} \textbf{\bibinfo{volume}{75}},
  \bibinfo{pages}{3969} (\bibinfo{year}{1995}),
  \urlprefix\url{https://link.aps.org/doi/10.1103/PhysRevLett.75.3969}.

\bibitem[{\citenamefont{Bradley et~al.}(1995)\citenamefont{Bradley, Sackett,
  Tollett, and Hulet}}]{Bradley_PRL_1995}
\bibinfo{author}{\bibfnamefont{C.~C.} \bibnamefont{Bradley}},
  \bibinfo{author}{\bibfnamefont{C.~A.} \bibnamefont{Sackett}},
  \bibinfo{author}{\bibfnamefont{J.~J.} \bibnamefont{Tollett}},
  \bibnamefont{and} \bibinfo{author}{\bibfnamefont{R.~G.} \bibnamefont{Hulet}},
  \bibinfo{journal}{Phys. Rev. Lett.} \textbf{\bibinfo{volume}{75}},
  \bibinfo{pages}{1687} (\bibinfo{year}{1995}),
  \urlprefix\url{https://link.aps.org/doi/10.1103/PhysRevLett.75.1687}.

\bibitem[{\citenamefont{Bogolubov and Bogolubov~Jr}(2010)}]{Bogolubov_2010}
\bibinfo{author}{\bibfnamefont{N.~N.} \bibnamefont{Bogolubov}}
  \bibnamefont{and} \bibinfo{author}{\bibfnamefont{N.~N.}
  \bibnamefont{Bogolubov~Jr}}, \emph{\bibinfo{title}{Introduction to quantum
  statistical mechanics}} (\bibinfo{publisher}{World Scientific},
  \bibinfo{year}{2010}).

\bibitem[{\citenamefont{Akhiezer and Peletminskii}(1981)}]{Akhiezer_1981}
\bibinfo{author}{\bibfnamefont{A.~I.} \bibnamefont{Akhiezer}} \bibnamefont{and}
  \bibinfo{author}{\bibfnamefont{S.~V.} \bibnamefont{Peletminskii}},
  \emph{\bibinfo{title}{Methods of Statistical Physics}}, vol.
  \bibinfo{volume}{104} of \emph{\bibinfo{series}{International Series in
  Natural Philosophy}} (\bibinfo{publisher}{Pergamon Press, Oxford},
  \bibinfo{year}{1981}),
  \urlprefix\url{http://www.sciencedirect.com/science/article/pii/B9780080250403500080}.

\bibitem[{\citenamefont{Girardeau and Arnowitt}(1959)}]{Girardeau_PhysRev_1959}
\bibinfo{author}{\bibfnamefont{M.}~\bibnamefont{Girardeau}} \bibnamefont{and}
  \bibinfo{author}{\bibfnamefont{R.}~\bibnamefont{Arnowitt}},
  \bibinfo{journal}{Phys. Rev.} \textbf{\bibinfo{volume}{113}},
  \bibinfo{pages}{755} (\bibinfo{year}{1959}),
  \urlprefix\url{https://link.aps.org/doi/10.1103/PhysRev.113.755}.

\bibitem[{\citenamefont{Valatin}(1961)}]{Valatin_PhysRev_1961}
\bibinfo{author}{\bibfnamefont{J.~G.} \bibnamefont{Valatin}},
  \bibinfo{journal}{Phys. Rev.} \textbf{\bibinfo{volume}{122}},
  \bibinfo{pages}{1012} (\bibinfo{year}{1961}),
  \urlprefix\url{https://link.aps.org/doi/10.1103/PhysRev.122.1012}.

\bibitem[{\citenamefont{{Nozières, P.} and {Saint James,
  D.}}(1982)}]{Nozieres_JPhysFr_1982}
\bibinfo{author}{\bibnamefont{{Nozières, P.}}} \bibnamefont{and}
  \bibinfo{author}{\bibnamefont{{Saint James, D.}}}, \bibinfo{journal}{J. Phys.
  France} \textbf{\bibinfo{volume}{43}}, \bibinfo{pages}{1133}
  (\bibinfo{year}{1982}),
  \urlprefix\url{https://doi.org/10.1051/jphys:019820043070113300}.

\bibitem[{\citenamefont{Shi and Griffin}(1998)}]{ShiPhysRep1998}
\bibinfo{author}{\bibfnamefont{H.}~\bibnamefont{Shi}} \bibnamefont{and}
  \bibinfo{author}{\bibfnamefont{A.}~\bibnamefont{Griffin}},
  \bibinfo{journal}{Phys. Rep.} \textbf{\bibinfo{volume}{304}},
  \bibinfo{pages}{1} (\bibinfo{year}{1998}),
  \urlprefix\url{https://www.sciencedirect.com/science/article/pii/S0370157398000155}.

\bibitem[{\citenamefont{Morgan}(2000)}]{MorganJPhysB2000}
\bibinfo{author}{\bibfnamefont{S.~A.} \bibnamefont{Morgan}},
  \bibinfo{journal}{J. Phys. B: At. Mol. Opt. Phys.}
  \textbf{\bibinfo{volume}{33}}, \bibinfo{pages}{3847} (\bibinfo{year}{2000}),
  \urlprefix\url{https://dx.doi.org/10.1088/0953-4075/33/19/303}.

\bibitem[{\citenamefont{Zagrebnov and Bru}(2001)}]{Zagrebnov_PhysRep_2001}
\bibinfo{author}{\bibfnamefont{V.~A.} \bibnamefont{Zagrebnov}}
  \bibnamefont{and} \bibinfo{author}{\bibfnamefont{J.-B.} \bibnamefont{Bru}},
  \bibinfo{journal}{Phys. Rep.} \textbf{\bibinfo{volume}{350}},
  \bibinfo{pages}{291} (\bibinfo{year}{2001}),
  \urlprefix\url{https://www.sciencedirect.com/science/article/pii/S0370157300001320}.

\bibitem[{\citenamefont{Proukakis and Jackson}(2008)}]{Proukakis_2008}
\bibinfo{author}{\bibfnamefont{N.~P.} \bibnamefont{Proukakis}}
  \bibnamefont{and} \bibinfo{author}{\bibfnamefont{B.}~\bibnamefont{Jackson}},
  \bibinfo{journal}{J. Phys. B: At. Mol. Opt. Phys.}
  \textbf{\bibinfo{volume}{41}}, \bibinfo{pages}{203002}
  (\bibinfo{year}{2008}),
  \urlprefix\url{https://dx.doi.org/10.1088/0953-4075/41/20/203002}.

\bibitem[{\citenamefont{Andersen}(2004)}]{Andersen_RevModPhys_2004}
\bibinfo{author}{\bibfnamefont{J.~O.} \bibnamefont{Andersen}},
  \bibinfo{journal}{Rev. Mod. Phys.} \textbf{\bibinfo{volume}{76}},
  \bibinfo{pages}{599} (\bibinfo{year}{2004}),
  \urlprefix\url{https://link.aps.org/doi/10.1103/RevModPhys.76.599}.

\bibitem[{\citenamefont{Popov}(1983)}]{Popov_1983}
\bibinfo{author}{\bibfnamefont{V.~N.} \bibnamefont{Popov}},
  \emph{\bibinfo{title}{Functional integrals in quantum field theory and
  statistical physics}} (\bibinfo{publisher}{D. Reidel, Dordrecht},
  \bibinfo{year}{1983}).

\bibitem[{\citenamefont{Popov}(1987)}]{Popov_1987}
\bibinfo{author}{\bibfnamefont{V.~N.} \bibnamefont{Popov}},
  \emph{\bibinfo{title}{Functional integrals and collective excitations}}
  (\bibinfo{publisher}{Cambridge University Press}, \bibinfo{year}{1987}).

\bibitem[{\citenamefont{Krasil'nikov and
  Peletminskii}(1993)}]{Krasilnikov_PhysParNucl_1993}
\bibinfo{author}{\bibfnamefont{V.~V.} \bibnamefont{Krasil'nikov}}
  \bibnamefont{and} \bibinfo{author}{\bibfnamefont{S.~V.}
  \bibnamefont{Peletminskii}}, \bibinfo{journal}{Phys. Part. Nucl.}
  \textbf{\bibinfo{volume}{24}} (\bibinfo{year}{1993}).

\bibitem[{\citenamefont{Akhiezer et~al.}(1994)\citenamefont{Akhiezer,
  Krasil'nikov, Peletminskii, and Yatsenko}}]{Akhiezer_PhysRep_1994}
\bibinfo{author}{\bibfnamefont{A.~I.} \bibnamefont{Akhiezer}},
  \bibinfo{author}{\bibfnamefont{V.~V.} \bibnamefont{Krasil'nikov}},
  \bibinfo{author}{\bibfnamefont{S.~V.} \bibnamefont{Peletminskii}},
  \bibnamefont{and} \bibinfo{author}{\bibfnamefont{A.~A.}
  \bibnamefont{Yatsenko}}, \bibinfo{journal}{Phys. Rep.}
  \textbf{\bibinfo{volume}{245}}, \bibinfo{pages}{1} (\bibinfo{year}{1994}),
  \urlprefix\url{https://www.sciencedirect.com/science/article/pii/0370157394900604}.

\bibitem[{\citenamefont{Griffin}(1996)}]{Griffin_PRB_1996}
\bibinfo{author}{\bibfnamefont{A.}~\bibnamefont{Griffin}},
  \bibinfo{journal}{Phys. Rev. B} \textbf{\bibinfo{volume}{53}},
  \bibinfo{pages}{9341} (\bibinfo{year}{1996}),
  \urlprefix\url{https://link.aps.org/doi/10.1103/PhysRevB.53.9341}.

\bibitem[{\citenamefont{Poluéktov}(2002)}]{Poluektov_LTP_2002}
\bibinfo{author}{\bibfnamefont{Y.~M.} \bibnamefont{Poluéktov}},
  \bibinfo{journal}{Low Temp. Phys.} \textbf{\bibinfo{volume}{28}},
  \bibinfo{pages}{429} (\bibinfo{year}{2002}),
  \urlprefix\url{https://doi.org/10.1063/1.1491184}.

\bibitem[{\citenamefont{Caballero-Benítez
  et~al.}(2013)\citenamefont{Caballero-Benítez, Paredes, and
  Romero-Rochín}}]{Romero-Rochin_2013_PLA}
\bibinfo{author}{\bibfnamefont{S.~F.} \bibnamefont{Caballero-Benítez}},
  \bibinfo{author}{\bibfnamefont{R.}~\bibnamefont{Paredes}}, \bibnamefont{and}
  \bibinfo{author}{\bibfnamefont{V.}~\bibnamefont{Romero-Rochín}},
  \bibinfo{journal}{Phys. Lett. A} \textbf{\bibinfo{volume}{377}},
  \bibinfo{pages}{1756} (\bibinfo{year}{2013}),
  \urlprefix\url{https://www.sciencedirect.com/science/article/pii/S0375960113004532}.

\bibitem[{\citenamefont{Bulakhov et~al.}(2018)\citenamefont{Bulakhov,
  Peletminskii, Peletminskii, Slyusarenko, and Sotnikov}}]{Bulakhov2018}
\bibinfo{author}{\bibfnamefont{M.~S.} \bibnamefont{Bulakhov}},
  \bibinfo{author}{\bibfnamefont{A.~S.} \bibnamefont{Peletminskii}},
  \bibinfo{author}{\bibfnamefont{S.~V.} \bibnamefont{Peletminskii}},
  \bibinfo{author}{\bibfnamefont{Y.~V.} \bibnamefont{Slyusarenko}},
  \bibnamefont{and} \bibinfo{author}{\bibfnamefont{A.~G.}
  \bibnamefont{Sotnikov}}, \bibinfo{journal}{J. Phys. B: At. Mol. Opt. Phys.}
  \textbf{\bibinfo{volume}{51}}, \bibinfo{pages}{205302}
  (\bibinfo{year}{2018}),
  \urlprefix\url{https://iopscience.iop.org/article/10.1088/1361-6455/aae061}.

\bibitem[{\citenamefont{Mordini et~al.}(2020)\citenamefont{Mordini,
  Trypogeorgos, Farolfi, Wolswijk, Stringari, Lamporesi, and
  Ferrari}}]{Mordini_2020_PRL}
\bibinfo{author}{\bibfnamefont{C.}~\bibnamefont{Mordini}},
  \bibinfo{author}{\bibfnamefont{D.}~\bibnamefont{Trypogeorgos}},
  \bibinfo{author}{\bibfnamefont{A.}~\bibnamefont{Farolfi}},
  \bibinfo{author}{\bibfnamefont{L.}~\bibnamefont{Wolswijk}},
  \bibinfo{author}{\bibfnamefont{S.}~\bibnamefont{Stringari}},
  \bibinfo{author}{\bibfnamefont{G.}~\bibnamefont{Lamporesi}},
  \bibnamefont{and} \bibinfo{author}{\bibfnamefont{G.}~\bibnamefont{Ferrari}},
  \bibinfo{journal}{Phys. Rev. Lett.} \textbf{\bibinfo{volume}{125}},
  \bibinfo{pages}{150404} (\bibinfo{year}{2020}),
  \urlprefix\url{https://link.aps.org/doi/10.1103/PhysRevLett.125.150404}.

\bibitem[{\citenamefont{Ota and Giorgini}(2020)}]{Ota_2020_PRA}
\bibinfo{author}{\bibfnamefont{M.}~\bibnamefont{Ota}} \bibnamefont{and}
  \bibinfo{author}{\bibfnamefont{S.}~\bibnamefont{Giorgini}},
  \bibinfo{journal}{Phys. Rev. A} \textbf{\bibinfo{volume}{102}},
  \bibinfo{pages}{063303} (\bibinfo{year}{2020}),
  \urlprefix\url{https://link.aps.org/doi/10.1103/PhysRevA.102.063303}.

\bibitem[{\citenamefont{Vianello and
  Salasnich}(2024)}]{Vianello_2024_ScientRep}
\bibinfo{author}{\bibfnamefont{C.}~\bibnamefont{Vianello}} \bibnamefont{and}
  \bibinfo{author}{\bibfnamefont{L.}~\bibnamefont{Salasnich}},
  \bibinfo{journal}{Sci. Rep.} \textbf{\bibinfo{volume}{14}},
  \bibinfo{pages}{15034} (\bibinfo{year}{2024}),
  \urlprefix\url{https://doi.org/10.1038/s41598-024-65897-2}.

\bibitem[{\citenamefont{Abrikosov et~al.}(1963)\citenamefont{Abrikosov, Gorkov,
  and Dzyaloshinski}}]{Abrikosov_1963}
\bibinfo{author}{\bibfnamefont{A.~A.} \bibnamefont{Abrikosov}},
  \bibinfo{author}{\bibfnamefont{L.~P.} \bibnamefont{Gorkov}},
  \bibnamefont{and} \bibinfo{author}{\bibfnamefont{I.~E.}
  \bibnamefont{Dzyaloshinski}}, \emph{\bibinfo{title}{Methods of quantum field
  theory in statistical physics}} (\bibinfo{publisher}{Prentice-Hall},
  \bibinfo{year}{1963}).

\bibitem[{\citenamefont{Timmermans et~al.}(1999)\citenamefont{Timmermans,
  Tommasini, Hussein, and Kerman}}]{Timmermans_PhysRep_1999}
\bibinfo{author}{\bibfnamefont{E.}~\bibnamefont{Timmermans}},
  \bibinfo{author}{\bibfnamefont{P.}~\bibnamefont{Tommasini}},
  \bibinfo{author}{\bibfnamefont{M.}~\bibnamefont{Hussein}}, \bibnamefont{and}
  \bibinfo{author}{\bibfnamefont{A.}~\bibnamefont{Kerman}},
  \bibinfo{journal}{Phys. Rep.} \textbf{\bibinfo{volume}{315}},
  \bibinfo{pages}{199} (\bibinfo{year}{1999}),
  \urlprefix\url{https://www.sciencedirect.com/science/article/pii/S0370157399000253}.

\bibitem[{\citenamefont{Radzihovsky et~al.}(2008)\citenamefont{Radzihovsky,
  Weichman, and Park}}]{Radzihovsky_AnnPhys_2008}
\bibinfo{author}{\bibfnamefont{L.}~\bibnamefont{Radzihovsky}},
  \bibinfo{author}{\bibfnamefont{P.~B.} \bibnamefont{Weichman}},
  \bibnamefont{and} \bibinfo{author}{\bibfnamefont{J.~I.} \bibnamefont{Park}},
  \bibinfo{journal}{Ann. Phys.} \textbf{\bibinfo{volume}{323}},
  \bibinfo{pages}{2376} (\bibinfo{year}{2008}),
  \urlprefix\url{https://www.sciencedirect.com/science/article/pii/S0003491608000742}.

\bibitem[{\citenamefont{Peletminskii et~al.}(2014)\citenamefont{Peletminskii,
  Peletminskii, and Poluektov}}]{Peletminskii_LTP_2014}
\bibinfo{author}{\bibfnamefont{A.~S.} \bibnamefont{Peletminskii}},
  \bibinfo{author}{\bibfnamefont{S.~V.} \bibnamefont{Peletminskii}},
  \bibnamefont{and} \bibinfo{author}{\bibfnamefont{Y.~M.}
  \bibnamefont{Poluektov}}, \bibinfo{journal}{Low Temp. Phys.}
  \textbf{\bibinfo{volume}{40}}, \bibinfo{pages}{500} (\bibinfo{year}{2014}),
  \urlprefix\url{https://doi.org/10.1063/1.4883893}.

\bibitem[{\citenamefont{Peletminskii et~al.}(2017)\citenamefont{Peletminskii,
  Peletminskii, and Slyusarenko}}]{Peletminskii_JPhysB_2017}
\bibinfo{author}{\bibfnamefont{A.~S.} \bibnamefont{Peletminskii}},
  \bibinfo{author}{\bibfnamefont{S.~V.} \bibnamefont{Peletminskii}},
  \bibnamefont{and} \bibinfo{author}{\bibfnamefont{Y.~V.}
  \bibnamefont{Slyusarenko}}, \bibinfo{journal}{J. Phys. B: At. Mol. Opt.
  Phys.} \textbf{\bibinfo{volume}{50}}, \bibinfo{pages}{145301}
  (\bibinfo{year}{2017}),
  \urlprefix\url{https://dx.doi.org/10.1088/1361-6455/aa75d6}.

\bibitem[{\citenamefont{Bloch et~al.}(2008)\citenamefont{Bloch, Dalibard, and
  Zwerger}}]{Zwerger_RevModPhys_2008}
\bibinfo{author}{\bibfnamefont{I.}~\bibnamefont{Bloch}},
  \bibinfo{author}{\bibfnamefont{J.}~\bibnamefont{Dalibard}}, \bibnamefont{and}
  \bibinfo{author}{\bibfnamefont{W.}~\bibnamefont{Zwerger}},
  \bibinfo{journal}{Rev. Mod. Phys.} \textbf{\bibinfo{volume}{80}},
  \bibinfo{pages}{885} (\bibinfo{year}{2008}),
  \urlprefix\url{https://link.aps.org/doi/10.1103/RevModPhys.80.885}.

\bibitem[{\citenamefont{Zwerger}(2011)}]{Zwerger_2011}
\bibinfo{author}{\bibfnamefont{W.}~\bibnamefont{Zwerger}},
  \emph{\bibinfo{title}{The BCS-BEC crossover and the unitary Fermi gas}}, vol.
  \bibinfo{volume}{836} (\bibinfo{publisher}{Springer Science \& Business
  Media}, \bibinfo{year}{2011}).

\bibitem[{\citenamefont{Chen et~al.}(2024)\citenamefont{Chen, Wang, Boyack,
  Yang, and Levin}}]{Levin_RevModPhys_2024}
\bibinfo{author}{\bibfnamefont{Q.}~\bibnamefont{Chen}},
  \bibinfo{author}{\bibfnamefont{Z.}~\bibnamefont{Wang}},
  \bibinfo{author}{\bibfnamefont{R.}~\bibnamefont{Boyack}},
  \bibinfo{author}{\bibfnamefont{S.}~\bibnamefont{Yang}}, \bibnamefont{and}
  \bibinfo{author}{\bibfnamefont{K.}~\bibnamefont{Levin}},
  \bibinfo{journal}{Rev. Mod. Phys.} \textbf{\bibinfo{volume}{96}},
  \bibinfo{pages}{025002} (\bibinfo{year}{2024}),
  \urlprefix\url{https://link.aps.org/doi/10.1103/RevModPhys.96.025002}.

\bibitem[{\citenamefont{Krasil'nikov et~al.}(1990)\citenamefont{Krasil'nikov,
  Peletminskii, and Yatsenko}}]{Krasilnikov_PhysA_1990}
\bibinfo{author}{\bibfnamefont{V.}~\bibnamefont{Krasil'nikov}},
  \bibinfo{author}{\bibfnamefont{S.}~\bibnamefont{Peletminskii}},
  \bibnamefont{and} \bibinfo{author}{\bibfnamefont{A.}~\bibnamefont{Yatsenko}},
  \bibinfo{journal}{Physica A} \textbf{\bibinfo{volume}{162}},
  \bibinfo{pages}{513} (\bibinfo{year}{1990}),
  \urlprefix\url{https://www.sciencedirect.com/science/article/pii/037843719090432R}.

\bibitem[{\citenamefont{Tolmachev}(1969)}]{Tolmachev_1969}
\bibinfo{author}{\bibfnamefont{V.~V.} \bibnamefont{Tolmachev}},
  \emph{\bibinfo{title}{Theory of a Bose gas}} (\bibinfo{publisher}{Moscow
  university press, in Russian}, \bibinfo{year}{1969}).

\bibitem[{\citenamefont{Peletminskii and
  Peletminskii}(2010)}]{Peletminskii_2010}
\bibinfo{author}{\bibfnamefont{A.~S.} \bibnamefont{Peletminskii}}
  \bibnamefont{and} \bibinfo{author}{\bibfnamefont{S.~V.}
  \bibnamefont{Peletminskii}}, \bibinfo{journal}{Low Temp. Phys.}
  \textbf{\bibinfo{volume}{36}}, \bibinfo{pages}{693} (\bibinfo{year}{2010}),
  \urlprefix\url{https://doi.org/10.1063/1.3490834}.

\bibitem[{\citenamefont{Peletminskii et~al.}(2013)\citenamefont{Peletminskii,
  Peletminskii, and Poluektov}}]{Poluektov_CMP_2013}
\bibinfo{author}{\bibfnamefont{A.~S.} \bibnamefont{Peletminskii}},
  \bibinfo{author}{\bibfnamefont{S.~V.} \bibnamefont{Peletminskii}},
  \bibnamefont{and} \bibinfo{author}{\bibfnamefont{Y.~M.}
  \bibnamefont{Poluektov}}, \bibinfo{journal}{Condens. Matter Phys.}
  \textbf{\bibinfo{volume}{16}}, \bibinfo{pages}{13603} (\bibinfo{year}{2013}),
  \urlprefix\url{https://icmp.lviv.ua/journal/zbirnyk.73/13603/abstract.html}.

\bibitem[{\citenamefont{Hugenholtz and Pines}(1959)}]{Hugenholtz_PhysRev_1959}
\bibinfo{author}{\bibfnamefont{N.~M.} \bibnamefont{Hugenholtz}}
  \bibnamefont{and} \bibinfo{author}{\bibfnamefont{D.}~\bibnamefont{Pines}},
  \bibinfo{journal}{Phys. Rev.} \textbf{\bibinfo{volume}{116}},
  \bibinfo{pages}{489} (\bibinfo{year}{1959}),
  \urlprefix\url{https://link.aps.org/doi/10.1103/PhysRev.116.489}.

\bibitem[{\citenamefont{Huang}(1987)}]{Huang_1987}
\bibinfo{author}{\bibfnamefont{K.}~\bibnamefont{Huang}},
  \emph{\bibinfo{title}{Statistical mechanics}} (\bibinfo{publisher}{John Wiley
  \& Sons}, \bibinfo{year}{1987}).

\bibitem[{\citenamefont{Bulakhov et~al.}(2021)\citenamefont{Bulakhov,
  Peletminskii, Slyusarenko, and Sotnikov}}]{Bulakhov_PS_2021}
\bibinfo{author}{\bibfnamefont{M.~S.} \bibnamefont{Bulakhov}},
  \bibinfo{author}{\bibfnamefont{A.~S.} \bibnamefont{Peletminskii}},
  \bibinfo{author}{\bibfnamefont{Y.~V.} \bibnamefont{Slyusarenko}},
  \bibnamefont{and} \bibinfo{author}{\bibfnamefont{A.~G.}
  \bibnamefont{Sotnikov}}, \bibinfo{journal}{Phys. Scr.}
  \textbf{\bibinfo{volume}{96}}, \bibinfo{pages}{045401}
  (\bibinfo{year}{2021}),
  \urlprefix\url{https://dx.doi.org/10.1088/1402-4896/abdcf5}.

\bibitem[{\citenamefont{Spada et~al.}(2022)\citenamefont{Spada, Pilati, and
  Giorgini}}]{Spada_PRA_2022}
\bibinfo{author}{\bibfnamefont{G.}~\bibnamefont{Spada}},
  \bibinfo{author}{\bibfnamefont{S.}~\bibnamefont{Pilati}}, \bibnamefont{and}
  \bibinfo{author}{\bibfnamefont{S.}~\bibnamefont{Giorgini}},
  \bibinfo{journal}{Phys. Rev. A} \textbf{\bibinfo{volume}{105}},
  \bibinfo{pages}{013325} (\bibinfo{year}{2022}),
  \urlprefix\url{https://link.aps.org/doi/10.1103/PhysRevA.105.013325}.

\bibitem[{\citenamefont{Prokof'ev et~al.}(2004)\citenamefont{Prokof'ev,
  Ruebenacker, and Svistunov}}]{Prokof'ev_PRA_2004}
\bibinfo{author}{\bibfnamefont{N.}~\bibnamefont{Prokof'ev}},
  \bibinfo{author}{\bibfnamefont{O.}~\bibnamefont{Ruebenacker}},
  \bibnamefont{and}
  \bibinfo{author}{\bibfnamefont{B.}~\bibnamefont{Svistunov}},
  \bibinfo{journal}{Phys. Rev. A} \textbf{\bibinfo{volume}{69}},
  \bibinfo{pages}{053625} (\bibinfo{year}{2004}),
  \urlprefix\url{https://link.aps.org/doi/10.1103/PhysRevA.69.053625}.

\bibitem[{\citenamefont{Arnold and Moore}(2001)}]{Arnold_PhysRevLett2001}
\bibinfo{author}{\bibfnamefont{P.}~\bibnamefont{Arnold}} \bibnamefont{and}
  \bibinfo{author}{\bibfnamefont{G.}~\bibnamefont{Moore}},
  \bibinfo{journal}{Phys. Rev. Lett.} \textbf{\bibinfo{volume}{87}},
  \bibinfo{pages}{120401} (\bibinfo{year}{2001}),
  \urlprefix\url{https://link.aps.org/doi/10.1103/PhysRevLett.87.120401}.

\bibitem[{\citenamefont{Nambu}(1960)}]{Nambu_PhysRev_1960}
\bibinfo{author}{\bibfnamefont{Y.}~\bibnamefont{Nambu}},
  \bibinfo{journal}{Phys. Rev.} \textbf{\bibinfo{volume}{117}},
  \bibinfo{pages}{648} (\bibinfo{year}{1960}),
  \urlprefix\url{https://link.aps.org/doi/10.1103/PhysRev.117.648}.

\bibitem[{\citenamefont{Goldstone}(1961)}]{Goldstone_NuovoCim_1961}
\bibinfo{author}{\bibfnamefont{J.}~\bibnamefont{Goldstone}},
  \bibinfo{journal}{Nuovo Cim. (1955-1965)} \textbf{\bibinfo{volume}{19}},
  \bibinfo{pages}{154} (\bibinfo{year}{1961}),
  \urlprefix\url{https://doi.org/10.1007/BF02812722}.

\bibitem[{\citenamefont{Gavoret and Nozières}(1964)}]{GavoretAnnPhys1964}
\bibinfo{author}{\bibfnamefont{J.}~\bibnamefont{Gavoret}} \bibnamefont{and}
  \bibinfo{author}{\bibfnamefont{P.}~\bibnamefont{Nozières}},
  \bibinfo{journal}{Ann. Phys.} \textbf{\bibinfo{volume}{28}},
  \bibinfo{pages}{349} (\bibinfo{year}{1964}),
  \urlprefix\url{https://www.sciencedirect.com/science/article/pii/0003491664902003}.

\bibitem[{\citenamefont{Kita}(2009)}]{Kita_PRB_2009}
\bibinfo{author}{\bibfnamefont{T.}~\bibnamefont{Kita}}, \bibinfo{journal}{Phys.
  Rev. B} \textbf{\bibinfo{volume}{80}}, \bibinfo{pages}{214502}
  (\bibinfo{year}{2009}),
  \urlprefix\url{https://link.aps.org/doi/10.1103/PhysRevB.80.214502}.

\bibitem[{\citenamefont{Kita}(2011)}]{Kita_JPhysSocJap_2011}
\bibinfo{author}{\bibfnamefont{T.}~\bibnamefont{Kita}}, \bibinfo{journal}{J.
  Phys. Soc. Jpn.} \textbf{\bibinfo{volume}{80}}, \bibinfo{pages}{084606}
  (\bibinfo{year}{2011}), \eprint{https://doi.org/10.1143/JPSJ.80.084606},
  \urlprefix\url{https://doi.org/10.1143/JPSJ.80.084606}.

\bibitem[{\citenamefont{S\"ut\ifmmode~\mbox{\H{o}}\else \H{o}\fi{} and
  Sz\'epfalusy}(2008)}]{Suto_PRA_2008}
\bibinfo{author}{\bibfnamefont{A.}~\bibnamefont{S\"ut\ifmmode~\mbox{\H{o}}\else
  \H{o}\fi{}}} \bibnamefont{and}
  \bibinfo{author}{\bibfnamefont{P.}~\bibnamefont{Sz\'epfalusy}},
  \bibinfo{journal}{Phys. Rev. A} \textbf{\bibinfo{volume}{77}},
  \bibinfo{pages}{023606} (\bibinfo{year}{2008}),
  \urlprefix\url{https://link.aps.org/doi/10.1103/PhysRevA.77.023606}.

\bibitem[{\citenamefont{Poluektov and Soroka}(2023)}]{Poluektov_JLTP_2023}
\bibinfo{author}{\bibfnamefont{Y.~M.} \bibnamefont{Poluektov}}
  \bibnamefont{and} \bibinfo{author}{\bibfnamefont{A.~A.}
  \bibnamefont{Soroka}}, \bibinfo{journal}{J. Low Temp. Phys.}
  \textbf{\bibinfo{volume}{210}}, \bibinfo{pages}{68} (\bibinfo{year}{2023}),
  \urlprefix\url{https://doi.org/10.1007/s10909-022-02885-8}.

\bibitem[{\citenamefont{Sosnick et~al.}(1989)\citenamefont{Sosnick, Snow,
  Sokol, and Silver}}]{Sosnick_EPL_1989}
\bibinfo{author}{\bibfnamefont{T.~R.} \bibnamefont{Sosnick}},
  \bibinfo{author}{\bibfnamefont{W.~M.} \bibnamefont{Snow}},
  \bibinfo{author}{\bibfnamefont{P.~E.} \bibnamefont{Sokol}}, \bibnamefont{and}
  \bibinfo{author}{\bibfnamefont{R.~N.} \bibnamefont{Silver}},
  \bibinfo{journal}{Europhys. Lett.} \textbf{\bibinfo{volume}{9}},
  \bibinfo{pages}{707} (\bibinfo{year}{1989}),
  \urlprefix\url{https://dx.doi.org/10.1209/0295-5075/9/7/016}.

\bibitem[{\citenamefont{Bogoyavlenskii
  et~al.}(1990)\citenamefont{Bogoyavlenskii, Karnatsevich, Kozlov, and
  Puchkov}}]{Bogoyavlenskii_LTP_1990}
\bibinfo{author}{\bibfnamefont{I.~V.} \bibnamefont{Bogoyavlenskii}},
  \bibinfo{author}{\bibfnamefont{L.~V.} \bibnamefont{Karnatsevich}},
  \bibinfo{author}{\bibfnamefont{Z.~A.} \bibnamefont{Kozlov}},
  \bibnamefont{and} \bibinfo{author}{\bibfnamefont{A.~V.}
  \bibnamefont{Puchkov}}, \bibinfo{journal}{Sov. J. Low Temp. Phys.}
  \textbf{\bibinfo{volume}{16}}, \bibinfo{pages}{77} (\bibinfo{year}{1990}),
  \eprint{https://pubs.aip.org/aip/ltp/article-pdf/16/2/77/20370762/77\_1\_10.0032339.pdf},
  \urlprefix\url{https://doi.org/10.1063/10.0032339}.

\bibitem[{\citenamefont{Moroni and Boninsegni}(2004)}]{Moroni_JLTP_2004}
\bibinfo{author}{\bibfnamefont{S.}~\bibnamefont{Moroni}} \bibnamefont{and}
  \bibinfo{author}{\bibfnamefont{M.}~\bibnamefont{Boninsegni}},
  \bibinfo{journal}{J. Low Temp. Phys.} \textbf{\bibinfo{volume}{136}},
  \bibinfo{pages}{129} (\bibinfo{year}{2004}),
  \urlprefix\url{https://doi.org/10.1023/B:JOLT.0000038518.10132.30}.

\bibitem[{\citenamefont{Glyde et~al.}(2011)\citenamefont{Glyde, Diallo, Azuah,
  Kirichek, and Taylor}}]{Glyde_PhysRevB_2011}
\bibinfo{author}{\bibfnamefont{H.~R.} \bibnamefont{Glyde}},
  \bibinfo{author}{\bibfnamefont{S.~O.} \bibnamefont{Diallo}},
  \bibinfo{author}{\bibfnamefont{R.~T.} \bibnamefont{Azuah}},
  \bibinfo{author}{\bibfnamefont{O.}~\bibnamefont{Kirichek}}, \bibnamefont{and}
  \bibinfo{author}{\bibfnamefont{J.~W.} \bibnamefont{Taylor}},
  \bibinfo{journal}{Phys. Rev. B} \textbf{\bibinfo{volume}{83}},
  \bibinfo{pages}{100507} (\bibinfo{year}{2011}),
  \urlprefix\url{https://link.aps.org/doi/10.1103/PhysRevB.83.100507}.

\bibitem[{\citenamefont{Gaunt et~al.}(2013)\citenamefont{Gaunt, Schmidutz,
  Gotlibovych, Smith, and Hadzibabic}}]{Gaunt_PhysRevA2013}
\bibinfo{author}{\bibfnamefont{A.~L.} \bibnamefont{Gaunt}},
  \bibinfo{author}{\bibfnamefont{T.~F.} \bibnamefont{Schmidutz}},
  \bibinfo{author}{\bibfnamefont{I.}~\bibnamefont{Gotlibovych}},
  \bibinfo{author}{\bibfnamefont{R.~P.} \bibnamefont{Smith}}, \bibnamefont{and}
  \bibinfo{author}{\bibfnamefont{Z.}~\bibnamefont{Hadzibabic}},
  \bibinfo{journal}{Phys. Rev. Lett.} \textbf{\bibinfo{volume}{110}},
  \bibinfo{pages}{200406} (\bibinfo{year}{2013}),
  \urlprefix\url{https://link.aps.org/doi/10.1103/PhysRevLett.110.200406}.

\bibitem[{\citenamefont{Rakhimov and Nishonov}(2025)}]{Rakhimov_PhysLettA2025}
\bibinfo{author}{\bibfnamefont{A.}~\bibnamefont{Rakhimov}} \bibnamefont{and}
  \bibinfo{author}{\bibfnamefont{M.}~\bibnamefont{Nishonov}},
  \bibinfo{journal}{Phys. Lett. A} \textbf{\bibinfo{volume}{531}},
  \bibinfo{pages}{130164} (\bibinfo{year}{2025}),
  \urlprefix\url{https://www.sciencedirect.com/science/article/pii/S0375960124008582}.

\end{thebibliography}
\end{document}